\documentclass{pinchcr}   

\title{Six out of equilibrium lectures}
\author{Jorge Kurchan}
\affiliation{ PMMH-ESPCI, CNRS UMR 7636, 10 rue Vauquelin, 75005 Paris, FRANCE}
\authors{2}

\begin{document}

\maketitle

\preface

The purpose of these notes is to introduce a group of subjects
in out of equilibrium statistical mechanics 
that have received considerable attention  in the
 last fifteen years or so.
They are mostly connected with
  time-reversibility and its relation to entropy,
are expressed in terms of large deviations, and
involve at some level the notion of timescale-separation.

We shall consider systems that contain, in their dynamic rules, an
element of noise.  There are good reasons for this: On the practical
side, a driven system needs to dissipate heat if it is to reach a
stationary regime rather than heating up indefinitely. Stochastic
systems, where energy is provided by the bath through time-dependent,
random forces, are well-studied physical models of heat reservoirs.  A
second, equally important reason is that very often, dynamical systems
in the presence of noise are much easier to study than purely
deterministic ones, because then very subtle ergodicity considerations
become trivial.  If the aim of ergodic theory is to understand how
randomness arises from deterministic constituents, once stochasticity
is added `by hand' the question is artificially bypassed. One may then
concentrate on the issues that are specific to non-equilibrium systems
with many degrees of freedom, just as  one postpones
ergodicity questions in the day to day practice of  equilibrium statistical mechanics.
One last consideration is that even purely deterministic systems are
sometimes more clearly understood as a small-noise limit: This
stochastic stability approach is very natural and appealing, not only
from the physical, but also from the mathematical point of view
(see e.g. \cite{lsy}).

Out of equilibrium statistical mechanics is a domain shared between
theoretical physicists, mathematical physicists and probabilists, a
fact reflected by a severely fragmented literature~
\footnote{I have followed, as general references, the following: the
books by Risken~\cite{Risken}, Gardiner~\cite{Gardiner} and Van
Kampen~\cite{Vankampen} for the general stochastic context.  Parisi's
book ~\cite{Parisi} in several places, in particular the relation
between sochastic and quantum mechanics.  Zinn-Justin's  book
~\cite{Zinn} for technical background on path integrals and a more
field-theoretic point of view.  I have also found very illuminating
the Lecture notes of H. Hilhorst~\cite{Hilhorst} (in French) and of
J. Cardy~\cite{Cardy}, where the reader may bridge
the main gap of these lectures: renormalisation.  The review of H\"anggi et
al~\cite{Hanggi} has a very comprehensive view of activation
processes.}  
Workers in each one of these fields have in mind a different network
of relations between subjects and techniques.  Two ideas that look
similar to a physicist may seem very distant to a probabilist, and
vice-versa.  The physicist's point of view -- the one I adopt --
stresses the relations between dynamics and the statistical mechanics
in space-time, between stochastic evolution and quantum mechanics, and
is on the alert for hidden symmetries and for scaling.

The first lecture  introduces stochastic dynamics in a formalism that
uses as much as possible the analogy with quantum mechanics, on the
assumption that the reader is already familiar with Shr\"odinger's equation.
Next, we discuss the consequences of 
 time-reversibility (in particular detailed balance), and how this is 
intimately related to thermodynamic equilibrium. Crucial for these
notes is the fact that the term responsible
for the breaking of 
a time-reversal symmetry in an out of equilibrium system 
is directly related to entropy production. 
In lecture three we discuss timescale separation and metastability.
In particular, we present through an example a general formalism for
 metastability~\cite{larry,bovier}
 based on the spectral decomposition of the dynamical operator. It is
quite elementary and intuitive, and is unjustifiably little known.  We
also describe very briefly the hydrodynamic limit, mainly to present
an example where fluctuations become weak through coarse-graining,
rather than through low temperatures.  This allows us to carry over the
`low noise' results, that we introduce for simplicity in the case of
low temperatures, for smooth, large-scale fluctuations: this is the
{\em Macroscopic Fluctuation Theory}~\cite{Bertini1}.  In lecture five we present two
forms of large-deviations: {\em i)} Low-Noise:  we stress as much as
possible the complete analogy between the
 `Freidlin-Wentzell'~\cite{Freidlin,low_noise}
theory and the standard WKB semiclassical approach of quantum
mechanics. {\em ii)} Deviations of  long-time averages: In the context of
glasses this is refrerred to as {\em Space-time Thermodynamics}~\cite{ChGa1}), 
because the long-time large deviation functions are in
complete analogy with $(d+1)$- dimensional thermodynamics. We shall
show that the phase transitions encountered in systems with slow dynamics
 within this formalism are closely related to
the spectral manifestations of metastability discussed in lecture
three.  With all these elements in hand, and playing with the
time-reversal symmetry and its breaking, we obtain  the
Fluctuation Theorem and Jarzynski's equality, the subject of the last
lecture.

The main aim  of this short course is to stimulate
 curiosity.
If it feels incomplete I will judge it successful.

\acknowledgements

I wish to thank J. Tailleur for reading of manuscript and suggestions.

\tableofcontents

\maintext

\chapter{Trajectories, distributions and path integrals.}

In this section we introduce  equations of motion containing a deterministic part, and 
a stochastic thermal bath. Next, we make the passage from a description in terms of trajectories to
one  in terms of distributions. The evolution of `probability clouds' in space 
is formally very close to (imaginary time) quantum mechanics. We do our best to exploit 
this analogy as much as possible because it opens the way for the application of all the methods in
quantum mechanics and field theory.  

\section{From trajectories to distributions}

\subsection{Trajectories}

Let us start by considering a system satisfying Hamilton's equations.
In addition, we shall allow for the possibility of external forces
 that do not derive from a potential acting on the
system,  throughout this work we shall denote them
${\bf f}({\bf q})$.  Nonconservative forces do work, and tend to heat the system
up. If we wish that the system  eventually become stationary 
we need a thermal
bath to absorb energy, both theoretically and in practice. The
simplest thermostat is the Langevin bath, consisting of friction term
proportional to the velocity and white noise. The equations of motion
read:
\begin{equation}
\left\{
\begin{array}{ll}
\dot q_i &= \frac{\partial \mathcal H}{\partial p_i} = \mbox{\em
  ... unless otherwise stated here ...}  = \frac{p_i}{m} =v_i\\ ~\\ \dot
p_i &= -\frac{\partial \mathcal H}{\partial q_i} -
\underbrace{f_i({\bf q})}_{Forcing} + \underbrace{\eta_i(t) - \gamma
  p_i}_{Thermal\; Bath}
\end{array}
\right.
\label{lange1}
\end{equation}
we shall use throughout $v_i=p_i/m$.
If the thermal bath is itself in equilibrium, noise intensity and
friction coefficient are related by:
\begin{equation}
\langle \eta_i(t)\eta_j(t')\rangle = 2 \gamma T \delta_{ij} \delta(t-t')
\label{firstkind}
\end{equation}
which, as we shall see, allows the system to equilibrate at
temperature $T$ in the absence of forcing.  The parameter $\gamma$
measures the intensity of coupling to the bath. In an unforced system
it does not affect the stationary distribution -- the Gibbs-Boltzmann
distribution $\sim e^{-\beta H}$ -- but it does control the dynamics:
compare for example a system of particles interacting with a potential
$V$ in equilibrium with a medium at temperature $T$, when the medium is
made of air  or of honey (small and large $\gamma$, respectively).

Equation (\ref{lange1}) can be justified in a number of ways. In
section 2.2.1 we shall see how this can be done.

Let us now consider the overdamped case of large $\gamma$. We can
formally (and somewhat dangerously)  neglect the acceleration term as follows:
\begin{eqnarray}
m \ddot q_i + \gamma \dot q_i + \frac{\partial V} {\partial q_i} + f_i
         &=& \eta_i(t) \nonumber \\ \Downarrow
         \;\;\;\;\;\;\;\;\;\;\;\;\;\;\;\;& & \nonumber \\ \gamma
         \frac{dq_i}{dt} + \frac{\partial V}{\partial q_i} + f_i &=&
         \eta_i(t) \nonumber \\ \Downarrow \;\;\;\;
         \;\;\;\;\;\;\;\;\;\;\;\; & & \nonumber \\ \frac{dq_i}{d\tau }
         + \frac{\partial V}{\partial q_i} + f_i &=& \eta_i(\tau)
\label{lange2}
\end{eqnarray}
In the last step we have rescaled time as $ \gamma \tau=t $, \\ which
yields $\langle \eta_i(\tau)\eta_j(\tau')\rangle = 2 T \delta_{ij}
\delta(\tau-\tau')$.  A few things to note in the passage to the
overdamped equation (\ref{lange2}) are:
\begin{itemize}
\item We now have half the dynamic variables.
\item The velocity $\dot q_i$ is now discontinuous in time, as is the
  noise itself.  We have to be careful what we mean by (\ref{lange2}),
  for example if we are going to programme it in a
  computer. We adopt the {\em \^Ito convention}, which means,
for example in one dimension:
\begin{equation}
q(t+\delta t)-q(t) + \delta t \; (V'+f)(q(t))=(\delta t)^{1/2}
\eta(t) 
\label{ito}
\end{equation}
The simplicity comes from the fact that the force is evaluated in the
old time $t$, and does not anticipate the result at the new time
$(t+\delta t)$~\cite{Vankampen,Gardiner,Risken,Hilhorst}.
\item
As we shall see, {\em these ambiguities}~\footnote{Which are
the analogue of factor-ordering ambiguities in quantum mechanics}
 {\em disappear when we consider
  the evolution of distribution functions instead of the
  trajectories.}
\item
Because the velocity is discontinuous, so is the quantity
$\sum_i f_i \dot q_i$: neglecting inertia makes power become a subtle
business. Indeed, some derivations, in particular involving work,
become more transparent keeping inertia, which makes velocities
a smooth function of time.
\end{itemize}

\subsection{Distributions}

Let us now change point of view, and consider the system rather than
from the point of view of individual trajectories with particular
noise realisations, as a `probability cloud' evolving in space. The
passage from the former to the latter description can be done in
several ways, and it is instructive to see their relations.

Consider the Langevin equation (\ref{lange2}). We wish to obtain the
equation of motion of the probability $P({\bf q},t)$.  Let us first do
it separately for a process in the absence of forces, and for a
process of advection in the absence of noise.  We obtain respectively:
\[
\begin{array}{ccccc}
 \dot q_i =&\eta_i \;\;\;\;\; &\longrightarrow& \;\;\;\;\;
 \frac{dP}{dt} &= T \nabla^2 P \\ \; & \; & \; & \; & \; \\ \dot q_i
 =&-\left(\frac{\partial V}{\partial q_i}+f_i\right) \;\;\;\;\;
 &\longrightarrow& \;\;\;\;\; \frac{dP}{dt} &= \sum_i \frac{\partial
 }{\partial q_i} \left[ \frac{\partial V}{\partial q_i}+f_i\right]P
\end{array}
\]
The first equation is just diffusion. The second uses the familiar
fact that if a distribution is carried by a flow $\dot q_i = g_i({\bf q})$,
its evolution is given by the 
 advective derivative $\dot P= \sum_i \frac {\partial (g_i
  P) } {\partial q_i}$.

{\em A useful remark:} A trick one implicitly uses often in
computer simulations is that whenever two processes act simultaneously
and their effect in a small time-interval is small, one gets the same
result by alternating short intervals with each acting alone. If the
separate evolutions are $\frac{dP}{dt}=-H_1 P$ and $\frac{dP}{dt}=-H_2
P$,  we can with the same argument recompose this as
$\frac{dP}{dt}=(H_1+H_2) P$. Applied to the previous situation, this
means that the evolution of the probability for the full Langevin
equation (\ref{lange2}) is:
\[
\begin{array}{ccccc}
 \dot q_i =&-\left(\frac{\partial V}{\partial q_i}+f_i\right) + \eta_i
\;\;\;\;\; &\longrightarrow& \;\;\;\;\; \frac{dP}{dt} &= \sum_i
\frac{\partial }{\partial q_i}\left[ T \frac{\partial }{\partial q_i}+
\frac{\partial V}{\partial q_i}+f_i\right]P \\ \; & \; & \; & \; & \;
\end{array}
\]
\begin{equation}
\dot P({\bf q},t) = -H_{FP}P({\bf q},t)
\label{fp}
\end{equation}
where we have defined the generator:
\begin{equation}
H_{FP}=-\sum_i \frac{\partial }{\partial q_i}\left[ T \frac{\partial
}{\partial q_i}+ \frac{\partial V}{\partial q_i}+f_i\right]
\label{Hfp}
\end{equation}
Writing (\ref{fp}) as a continuity equation, we identify the current:
\begin{equation}
\dot P = -{\mbox{div}} \,   {\bf J} \;\;\;  with \; the \; definition 
 \;\;\; J_i({\bf q})= \left[ T \frac{\partial }{\partial q_i}+
\frac{\partial V}{\partial q_i}+f_i\right]P
\end{equation}

Exactly the same procedure can be used to derive the evolution of the
probability for the case with inertia (\ref{lange1}).  In order to
split the system in a diffusion and an advection term, we need to work
in phase space. The result is the Kramers equation:
\begin{equation}
\dot P({\bf q},{\bf p},t) = -H_{K}P({\bf q},{\bf p},t)
\label{kr}
\end{equation}
with
\begin{equation}
H_K= \underbrace{ \frac{\partial {\mathcal H}}{\partial p_i}
\frac{\partial }{\partial q_i}- \frac{\partial {\mathcal H}}{\partial
q_i} \frac{\partial }{\partial p_i}
}_{H_{Liouville}~:~\{\;,\;\}} \;\;
\;\; \underbrace{- \gamma \frac{\partial }{\partial p_i}\left(T
\frac{\partial }{\partial p_i} + \frac{p_i}{m} \right)}_{H_b~:~bath} \;\;\;
\underbrace{ - \frac{\partial }{\partial p_i} f_i({\bf q})
}_{H_f~:~forcing}
\label{Hk}
\end{equation}
(summation convention), where we recognise the Poisson bracket
associated with Hamilton's equations, plus a bath, and (eventually) a
forcing term.

Again, writing the Kramers equation as a continuity equation:
\begin{equation}
  \frac{\partial P({\bf q,p},t)}{\partial t}=-H_K P({\bf
    q,p},t)=-\mbox{div} {\bf J} = - \sum_i \left(\frac{\partial
    J_{q_i}}{\partial q_i} + \frac{\partial J_{p_i}}{\partial p_i}
    \right),
\label{Kramers1}
\end{equation}
we identify a   current:
\begin{equation}
  J_{q_i}= \frac{\partial \mathcal{H}}{\partial p_i} P({\bf q,p},t) \quad
  \quad \quad \quad J_{p_i}= 
  - \left( \gamma T \frac{\partial }{\partial p_i}+\gamma \frac{\partial
      \mathcal{H}}{\partial p_i} 
    +\frac{\partial \mathcal{H}}{\partial q_i} + f_i\right)P({\bf q,p},t).
\label{current}
\end{equation}
A surprise is that, even in an unforced, ${\bf f}=0$ equilibrium
stationary state the phase-space current is nonzero. This suggests
that we define an alternative quantity, the {\em reduced phase-space
  current} as~\cite{sorin}:
\begin{equation}
  J^{{red}}_{q_i}\equiv J_{q_i} +T \frac{\partial P({\bf
  q,p})}{\partial p_i} \quad \quad \quad \quad J^{{red}}_{p_i}= J_{p_i}- T
  \frac{\partial P({\bf q,p})}{\partial q_i}.
\label{reduced}
\end{equation}
The currents (\ref{current}) and (\ref{reduced}) differ by a term
without divergence, and hence their fluxes over closed surfaces
coincide. The interesting property of (\ref{reduced}) is that it is
zero for the canonical distribution, as one can easily
check. Furthermore, in a case with metastable states it is small
everywhere, and it is concentrated along reaction paths.

\vspace{.2cm}

\fbox{\parbox{12cm} {%

{\bf Gaussian thermostat}

\vspace{.1cm}

In some situations one wishes to study a system with thermostat that
preserves the energy.  This can be done with a deterministic
(noiseless) `Gaussian'~\cite{Hoover} thermostat, extensively used in
the context of entropy production and the Gallavotti-Cohen theorem.
As we shall see later, it is in some cases convenient to have in
addition a small amount of energy-preserving noise. We thus consider:
\begin{equation}
\left\{
\begin{array}{ll}
\dot q_i &= \frac{p_i}{m} \\ ~\\ \dot p_i &= -\frac{\partial {
 \mathcal H}}{\partial q_i} - g_{ij} (\eta_j - f_j({\bf q}))=
 -\frac{\partial { \mathcal H}}{\partial q_i} + \underbrace{ g_{ij}
 \eta_j}_{\stackrel{conservative}{noise}} - \underbrace{f_i({\bf
 q})}_{forcing} + \underbrace{\gamma(t) p_i}_{thermostat}
\end{array}
\right.
\label{Gau}
\end{equation}
\begin{itemize}
\item $\eta_j$ are white, independent noises of variance $\epsilon$, unrelated 
to temperature,
since the energy is fixed.
\item $g_{ij}=\delta_{ij}-\frac{p_i p_j}{{\bf p^2}}$ is the projector
onto the space tangential to the energy surface.
\item  Multiplying the first of (\ref{Gau}) by 
$\frac{\partial V({\bf q})}{\partial q_i}$, the second
by $\frac{p_i}{m}$ and adding, one concludes that energy is conserved
provided $\gamma(t) = \frac{{\bf f}\bullet {\bf p}}{{\bf p^2}}$.
\end{itemize}
 The product $g_{ij}({\bf p})\eta_j$ is rather ill-defined because both
$g_{ij}$  and $\eta_j$  are discontinuous functions of time. The ambiguity is
raised by discretising time~\cite{Risken}, or by specifying the evolution
of probability, as we now do.

Repeating the steps leading to the  Fokker-Planck and Kramers equation,
we find that the probability evolves through:
\begin{equation}
\dot P({\bf q},{\bf p})= -H_G P({\bf q},{\bf p}) \label{eq2}
\end{equation}
where $H$ is the operator:
\begin{equation}
 H_G= \frac{p_i}{m} \frac{\partial}{\partial q_i}-\frac{\partial V({\bf
q})}{\partial q_i} \frac{\partial}{\partial p_i}+
\frac{\partial}{\partial p_i}\; [\gamma p_i] -
\frac{\partial}{\partial p_i} f_i - \epsilon \frac{\partial}{\partial
p_j} g_{ij} g_{il} \frac{\partial}{\partial p_l}
\label{eq3}
\end{equation}
where the summation convention is assumed.
 The precise factor ordering in the last term is important,
and specifies the meaning of
Eq. (\ref{Gau}).
 In the absence of driving ${\bf f}=0$ it is easy to check that $H$
annihilates any function that depends on the phase-space coordinates 
only through the energy ${\mathcal H}=\frac{{\bf p}^2}{2m}+V $. 
Hence, the noise respects the microcanonical
 measure, in that case.
}}

\subsection{ Other spaces. Doi-Peliti variables}

The Hilbert spaces associated with probability distributions of the
Fokker-Planck and Kramers equations are different, as  the former 
consists of functions of $N$-dimensional space $P({\bf q})$ and the
latter of $2N$-dimensional space $P({\bf q},{\bf p})$.  In fact,
other spaces appear naturally~\cite{Peliti_Doi,Hilhorst}
 when the dynamic variables are not
continuous.  This does not bring in any new conceptual feature, but it
allows to write other stochastic problems in a familiar `quantum'
notation. Let us give two examples:

\vspace{.1cm}

{\bf Bosons}

\vspace{.1cm}

We consider particles on a lattice, with no exclusion. Denoting $n_i$
number of particles in site $i$, the complete set of configurations is
spanned by the space:
\begin{equation}
|n\rangle = |n_1,\ldots,n_N\rangle = \otimes_{i=1}^N |n_i\rangle\;,
\end{equation}
so that a a probability distribution is written as:
\begin{equation}
P= \sum_{n_1,...,n_N} c_{n_1,...,n_N} |n_1,\ldots,n_N\rangle
\end{equation}
We can write any evolution operator  in this space introducing the generators
\begin{eqnarray}
a_i  |n_i\rangle  &=&n_i |n_i-1\rangle \nonumber \\
a_i^\dag |n_i\rangle &=& |n_i+1\rangle \nonumber \\
a_i|0\rangle &=&0
\end{eqnarray}
Note that $a_i^\dag$ and $a_i$ are not mutually Hermitian conjugates.
For example, for simple diffusion on a one-dimensional lattice reads $\dot P = -HP$
with
\begin{equation}
H=-\sum_i  (a^\dag_{j+1} + a^\dag_{j-1} -2 a^\dag_j)a_j
\end{equation}

\vspace{.1cm}

{\bf Spins: the Simple Symmetric Exclusion Process}

\vspace{.1cm}

Boson variables do not lend themselves easily to processes where
 occupation of sites is limited, because particles exclude one another.
In those cases, Fermions and Spins, which have a finite Hilbert space, 
 appear naturally.
To be more definite consider the `Simple Symmetric Exclusion
Process' on a one-dimensional lattice.
This corresponds to the stochastic process on the lattice $\{1,\ldots,N\}$
where particles jump   to  neighbouring sites, but with the limitation
that each site
can accommodate at most one particle.
Configurations $n\in\{0,1\}^N$ are then identified with ket states
\begin{equation}
|n\rangle = |n_1,\ldots,n_N\rangle = \otimes_{i=1}^N |n_i\rangle\;,
\end{equation}
which specify the occupation number of each site, namely
$n_i\in\{0,1\}$.  The bulk evolution is given by
 the transition rates
\begin{eqnarray}
w(n^{i+1,i},n) & = & - \langle n^{i,i+1}|H_B|n\rangle
=(1 - n_i)n_{i+1}\nonumber \\ w(n^{i,i+1},n) & = & - \langle
n^{i,i+1}|H_B|n\rangle =n_i(1 - n_{i+1})\;.
\end{eqnarray}
where $n^{i,j}$ is the configuration which is obtained from the
configuration $n$ by removing a particle in $i$ and adding it in $j$.

We introduce the operators $S$, which act as
\begin{eqnarray}
S^+_i |n_i\rangle &=& (1-n_i) |n_i+1\rangle \nonumber \\
S^-_i |n_i\rangle &=& n_i      |n_i-1\rangle \nonumber \\
S^0_i |n_i\rangle &=& \left(n_i-\frac12\right)  |n_i\rangle\;.
\end{eqnarray}
and
satisfy the SU(2) algebra
\begin{eqnarray}
\label{commutatorsSU2}
[S_i^{0},S_i^{\pm}] &=& \pm S_i^{\pm} \nonumber \\
{[}S_{i}^{-},S_{i}^{+}{]} &=& -2S_i^{0}\;.
\end{eqnarray}
Note again that, in this representation, $S^{\pm}$ are not mutually
Hermitian conjugates.

In terms of these, the evolution  of the SSEP  is generated by a  spin one-half 
ferromagnet \cite{Schutz}
\begin{equation}
\label{bulksu2}
- H_B = \sum_{i=1}^{N-1} \left( S^{+}_iS^{-}_{i+1} +
S^{-}_iS^{+}_{i+1} + 2 S^{0}_iS^{0}_{i+1} - \frac12 \right)\;,
\label{SSEP}
\end{equation}

\vspace{.2cm}

\section{Hilbert Space}
We shall throughout these lectures use the bracket notation, following
Kadanoff and Swift~\cite{KS} Doing this, we uncover the similarities
and differences between stochastic and quantum dynamics, and allows us
to import many techniques developed in quantum many-body and field
theory.  Here we do everything for the Fokker-Planck case, the
generalisation to other dynamics is straightforward.  We define, as
usual, the ${\bf q}$ representation:
\begin{equation}
P({\bf q})= \langle {\bf q}| \psi \rangle
\end{equation}
The evolution equation becomes:
\begin{equation}
\frac{d}{dt}| \psi \rangle = - H_{FP} | \psi \rangle
\end{equation}
whose solution is 
\begin{equation}
 | \psi(t) \rangle =  e^{-t H_{FP}}| \psi_o \rangle  \;\;\; \rightarrow \;\;\;
P({\bf q},t)= \langle {\bf q}| e^{-t H_{FP}}| \psi_o \rangle
\end{equation}
The transition probability is given by the matrix element:
\begin{equation}
P({\bf q},t \; ;{\bf q_o},t_o ) = \langle {\bf q}| e^{-(t-t_o)
H_{FP}}|{\bf q_o} \rangle
\end{equation}

As in quantum mechanics, it is useful to consider the spectrum of
$H_{FP}$. Because $H_{FP}$ is not self-adjoint, we have to distinguish right and
left eigenvectors:
\begin{equation}
 H_{FP} | \psi_a^R \rangle = \lambda_a | \psi_a^R \rangle \;\;\;\; ;
\;\;\;\; \langle \psi_a^L | H_{FP} = \lambda_a \langle \psi_a^L|
\end{equation}
The resolution of the identity is:
\begin{equation}
\langle \psi_a^L | \psi_b^R \rangle = \delta_{ab} \;\;\;\; ; \;\;\;\;
 {\bf {I}}= \sum_a | \psi_a^R \rangle\langle \psi_a^L |
\end{equation}

Defining the (in general, unnormalisable) `flat' state $\langle - |$
as a constant over all configurations:
\begin{equation}
 \langle - |{\bf q}\rangle =1 
\end{equation}
and using the fact that conservation of probability implies that, at
all times
\begin{equation}
1= \sum_{all \; configurations}\langle {\bf q}|\psi (t) \rangle= \langle -
|\psi(t)\rangle \;\; \rightarrow \;\; \frac{d}{dt}\langle - |\psi(t)
\rangle= -\langle - |H_{FP} |\psi(t) \rangle=0 \;\;\; \forall \psi(t)
\end{equation}
and the fact that a stationary state satisfies
\begin{equation}
\frac{d}{dt} |{{ stat } }\rangle =- H_{FP} |{{ stat } }  \rangle=0
\end{equation}
we have that `flat' and stationary states are the left and right
zero-eigenvalue eigenvectors
\begin{equation}
   \langle - |H_{FP} =0 \;\;\; \;\;\; ; \;\;\; \;\;\; H_{FP}
   |{{ stat } } \rangle=0
\end{equation}
Finally, 
writing 
\begin{equation}
  | \psi_o \rangle = \sum_a \; c_a \; | \psi_a^R \rangle 
\end{equation}
we have
\begin{equation}
 P({\bf q},t) = \langle {\bf q} | \psi(t) \rangle = \sum_a \; c_a \;
\langle {\bf q}| \psi_a^R \rangle e^{-t \lambda_a}
\label{develop}
\end{equation}
Equation (\ref{develop}) already shows us that $\lambda_a$ cannot be
negative, and that the long-time properties of the probabilities are
encoded in the eigenvectors with small eigenvalues.
In particular, if all eigenvalues have  real parts larger than zero,
there is no stationary state': this happens typically if the system is unbounded.

In the Kramers case, all of what we have said applies, provided one considers functions:
\begin{equation}
P({\bf q},{\bf p})= \langle {\bf q},{\bf p} | \psi \rangle
\end{equation}

\vspace{.2cm}

\fbox{\parbox{12cm} {%
{\bf Perron-Frobenius theorem}: $P$  is a probability
distribution, so it has to be positive everywhere at all times.  If one 
$\lambda_a$ has a negative  real part, it  dominates the 
sum (\ref{develop}) for large times, its coefficient going to infinity.
  Because  $\langle -|\psi_a \rangle =0$, necessarily   $\langle {\bf q}|\psi_a \rangle $ takes
positive and  negative values, and  this will make  $P(t)$
at large times not everywhere positive, contrary to the assumption. }}

\vspace{.2cm}

\subsection{Correlations and responses}

Correlation functions are averages over
 many realizations of the process, with different
realisations of the random noise each time.
 They can be expressed in this notation as:
\begin{equation}
C_{AB}(t,t') = \langle A(t) B(t') \rangle_{realisation} = \langle - | A \;
e^{-(t-t')H} \; B \; e^{-t'H} | {{ init }} \rangle
\end{equation}
here $H$ may be the Fokker-Planck, Kramers, or in general any Doi-Peliti  operator.
Similarly for the response functions
\begin{equation}
R_{AB}(t,t') = \frac{\delta\langle A(t) \rangle}{\delta h_B(t')} =
 \langle - | A \; e^{-(t-t')H} \; \frac{dH}{dh_B} \; e^{-t'H} |
 {{ init }} \rangle
\end{equation}
where the average $\langle A \rangle$ is over noise realisations.
Eigenvalues with small real parts 
have something to say about the decay of long-time
correlations:
\begin{eqnarray}
C(t,t') &=&\langle A(t) A(t') \rangle = \langle -|A e^{-(t-t')H} A |init \rangle
 = \sum_a \langle -|A|\psi^R_a\rangle \langle\psi^L_a |A|init \rangle
e^{-\lambda_a t} \nonumber \\
&\sim&  \langle -|A|stat\rangle \langle - |A|init \rangle
+\langle -|A|\psi^R_1\rangle \langle\psi^L_1 |A|init \rangle
e^{-\lambda_{1} t}
\label{gap0}
\end{eqnarray}
which decays to the asymptotic 
value as an exponential of ${\mbox{Re}} \lambda_1$,
the first eigenvector with non-zero real part. The existence of
gap between lowest and first eigenvalue leads to exponential decays of 
correlations.

\subsection{Conserved quantities}

Consider a probability distribution  that is concentrated  on
an energy  shell, for example in phase space $  E({\bf q},{\bf p})=E_o$.
This can be expressed as an eigenvalue equation: 
\begin{equation}
 E({\bf q},{\bf p}) P({\bf q},{\bf p}, t) = E_o P({\bf q},{\bf p}, t)
\end{equation}
Consider further an evolution that conserves energy, generated by some
$H$.  By assumption, there are no transitions between shells of
different energies, so the matrix of $H$ is of block form, each one
corresponding to a value of $E_o$.  This in turn implies that $E$
commutes with the operator $H$:
\begin{equation}
\label{derivata-energia}
 [E, H] = 0\;.
\end{equation}
Clearly, what we have said applies to any conserved quantity.

The fact that the stochastic process can be now broken into subspaces of fixed energy
entails the existence for each $E_o$ of 
a different stationary state  
 $|stat_{E_o}\rangle$, and  the corresponding flat measure $\langle -_{E_o}|$.

\subsection{Analogy with quantum mechanics}

 Making the identification:
\begin{equation}
\hat q_i^{op} \longrightarrow T \frac{\partial}{\partial q_i}
\end{equation}
 we may write
\begin{equation}
\begin{array}{cclc}
\tilde H \equiv T H_{FP} =& -\underbrace{\left(T \frac{\partial}{\partial
 q_i} \right)}_{\downarrow}& &\left[ \underbrace{\left( T
 \frac{\partial}{\partial q_i}\right)}_{\downarrow} + \frac{\partial
 V}{\partial q_i} + f_i \right] \nonumber \\  =&- {\hat
 q_i}^{op}& &\left[ \;\; {\hat q_i}^{op} + \frac{\partial V}{\partial
 q_i} + f_i \right]
\end{array}
\end{equation}
and the evolution is
\begin{equation}
\dot P = \underbrace{-\frac{1}{T} \tilde H P}_{H_{FP}}
\label{class11}
\end{equation}

The situation is analogous to quantum mechanics, with $T$ playing the role of $\hbar$
and the $ {\hat q_i}^{op}$ the role of `momentum operators'. 
The quantum-like Hamiltonian comes from a `classical'   Hamiltonian 
\begin{equation}
\tilde H= - {\hat
 q_i}\left[ \;\; {\hat q_i} + \frac{\partial V}{\partial
 q_i} + f_i \right]
\label{class2}
\end{equation}
in a phase space with twice the number of variables.
 This `classical' problem with double number of degrees of freedom will  play a key role in the next sections.

\section{Functional approach}

Let us first consider a general and extremely powerful 
trick to convert an algebraic problem of finding
the roots of a system of  equations $Q_a( {\bf x})=0$
 into a statistical problem
of calculating a partition function. We compute the sum over roots 
of a function $A$:
\begin{equation}
\langle A({\bf x}) \rangle \equiv
\sum \left. A({\bf x}) \right|_{\mbox{roots  of } Q_a }
\end{equation}
 This can be written 
\begin{eqnarray}
\langle A( {\bf x}) \rangle &=& \int \; d{\bf x} \; A({\bf x})    \; 
\delta(Q_a({\bf x})) \;  {\mbox{det}} \left|\frac{\partial Q_c}{\partial x_b}({\bf x})\right| 
 \nonumber \\ &=&
 \int \; d{\bf x} \;  \int_{-i \infty}^{i \infty} \;
d  {\hat{{\bf x}}} \; A({\bf x})\; {\mbox{det}} 
\left|\frac{\partial Q_c}{\partial x_b}({\bf x})\right| 
  \; e^{ \sum_a \hat{x}_a Q_a({\bf x})}    
\end{eqnarray}
The determinant is there to insure that we count each root with a unit weight. We have thus obtained 
an expression that has the form of a partition function.
 
The Martin-Siggia-Rose/DeDominicis-Janssen~\cite{MSR} approach consists of
applying this technique to differential, rather than ordinary equations, in this case the Langevin
equation. 
To make this properly,
one should discretise time and write a delta function imposing the 
equation (\ref{ito}) at each time. 
One obtains a sum over paths:
\begin{equation}
P(q,t)=\overbrace{\int D[q]}^{\stackrel{paths \; with}{appropriate \; 
b.c's}} \; \; \underbrace{D[\eta]}_{noise} \;\; \overbrace{\delta\left[\dot
q+ \frac{\partial V}{\partial q} + \eta\right]}^{functional \; delta} \;
\; \underbrace{e^{-\int dt\; \eta^2/4T}}_{path \; probability}
\label{popo}
\end{equation}
Where for the moment we stay in   one dimension,
and  omit the forcing term, the generalisation to several dimensions and  ${\bf f}\neq 0$ is straightforward. 
The determinant  is unity in the Ito convention (\ref{ito}),
because the matrix of second derivatives (containing the derivative
of the equation of motion at time $t_i$ with respect to $q(t_j)$)
  is upper triangular  with  ones in the diagonal.
 
Introducing the integral representation of the delta with a `hat' variable
(one per time):
\begin{equation}
P(q,t)=\int D[q]  \; \; D[\eta] \;\; \int_{-i \infty}^{+ i \infty} 
\overbrace{D[\hat q]}^{`response' \; field} \; \;
e^{\int dt \; \left[{\hat q_i} \left(\dot
q+ \frac{\partial V}{\partial q} + \eta\right)\;- \eta^2/4T\right]} 
\end{equation} 
The boundaries are free for $\hat q$.
Integrating the Gaussian noise away, we obtain:
\begin{eqnarray}
P(q,t)=\int D[q] \; D[\hat q] & & \exp{\int dt \; \left[ \hat q \dot q
    - \underbrace{\left\{ - \hat q \left(T \hat q + \frac{\partial
        V}{\partial q} \right) \right\} } \right] } =\int D[q, \hat q]
\; \; e^{\pm \int dt \; [\hat q \dot q - H]} \nonumber \\ &
&\;\;\;\;\;\;\;\;\; \;\;\;\;\;\;\;\;\; \;\;\;\; \;\;\;\;
\;\;\;\;\;\;\;\;\; \stackrel{\downarrow}{ H}
\label{oooo}
\end{eqnarray}
Making the change of variables  $\hat q \rightarrow \frac{\hat q}{T} $ the evolution
equation becomes, for many degrees of freedom:
\begin{eqnarray}
P(q,t)=\int D[{\bf q}, {\bf \hat q}] & & \exp{\frac{1}{T} \int dt \;
\left[ \hat q_i \dot q_i - \underbrace{ \left\{ - \hat q_i \left( \hat
q_i + \frac{\partial V}{\partial q_i} \right) \right\} } \right] }
\nonumber \\ & & \;\;\;\;\;\;\;\;\; \;\;\;\;\;\;\;\;\; \;\;\;\;
\;\;\;\; \;\;\;\; \;\;\;\;\;\;\;\;\;\stackrel{\downarrow}{ \tilde H}
\label{eqq}
\end{eqnarray}
(summation convention). 
The  exponent is multiplied by $1/T$, which thus plays the role of $1/\hbar$.
From what we know of the path -integral representation of
 quantum mechanics, 
we recognise the action associated with the evolution equation (\ref{class11}).
The analogy with  (imaginary time) quantum mechanics can be carried further {\em in the absence of
forcing}, by making  the transformation:
\begin{equation}
H_h= e^{\beta V/2} \; H_{FP} \; e^{-\beta V/2} = \frac{2}{T} \sum_i
 \left[ - \frac{T^2}{2} \frac{\partial^2}{\partial q_i^2} +
 \frac{1}{8} \left(\frac{\partial V}{\partial
 q_i}\right)^2 - \frac{T}{4} \frac{\partial^2 V}{\partial
 q_i^2}\right]
\label{hermi1}
\end{equation}
We now have an $H_h$ which is truly of the Shr\"odinger form, albeit with
a new potential $V_{eff}=  \frac{1}{8} \left(\frac{\partial V}{\partial
 q_i}\right)^2 - \frac{T}{4} \frac{\partial^2 V}{\partial
 q_i^2}$

\begin{figure}[htbp]
\begin{center}
\includegraphics[width=7cm]{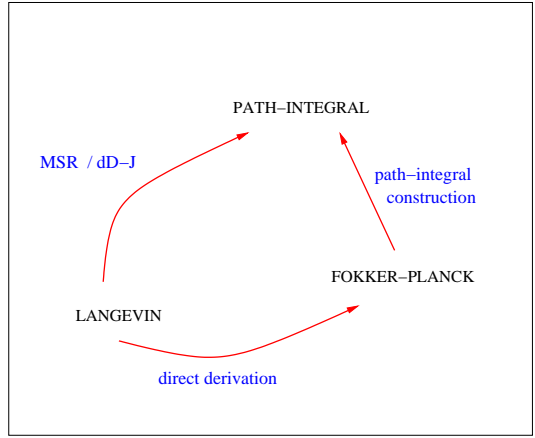}
\end{center}
\caption{With the MSR/dDJ one can go straight from stochastic equation
to the path-integral. Otherwise, given the equation of motion for the 
probability distribution
one can express it as a path integral as in quantum mechanics textbooks.}
\label{diagra}
\end{figure}

All in all, the situation is as in Fig. \ref{diagra}: we can go from
the Langevin equation to the Fokker-Planck equation for the
distribution directly, and then  construct a path integral representation of
the evolution using the analogy between Fokker-Planck and Shr\"odinger
equations. Alternatively, we can go straight from the Langevin
equation to the path integral, through the
Martin-Siggia-Rose/deDominicis-Janssen construction.  The latter
procedure is much more flexible, and is thus extensively
 used in the physics literature.

\subsection{Lagrangian (Onsager-Machlup) form}

As in all phase-space problems where the momenta appear quadratically
(in this case the `hat' variables $\hat q_i$), we have the possibility
of integrating them away, thus going to the Lagrangian
representation~\cite{Onsager}. Alternatively, we can go straight to this
representation by integrating (\ref{popo}) over the noise {\em before}
introducing the `hat' variables.  The result is:
\begin{equation}
P(q,t)=\int D[{\bf q}] \; \exp{\left[ - \frac{1}{4T}\int dt \;
    \left(\dot q_i + \frac{\partial V}{\partial q_i}+ f_i \right)^2
    \right]}
\label{lag}
\end{equation}
Note that we have inherited the \^Ito convention from (\ref{ito}),
which means a definite prescription on how to discretise the time
integral.  This sum takes the form of a partition
function at temperature $\sim T$. This means we can in principle
compute it using well-established Monte Carlo methods: {\em a nonequilibrium problem
for the configurations becomes an `equilibrium' problem in trajectory-space}.   

In the absence of forcing ${\bf f}=0$ we can develop the square in the
exponent, and recognising that the double product is a total
derivative, we have:
\begin{equation}
P(q,t)=\int D[{\bf q}] \;
\exp
\left\{
- \frac{1}{4T}\int dt  \;\left[
\dot q_i^2 +
\left(  \frac{\partial V}{\partial q_i}     \right)^2   -2T  \left(  \frac{\partial^2 V}{\partial q_i^2}     \right) \right]
\right\} \;
e^{- \frac{1}{2T}[V({\bf q})-V({\bf q_o})]}
\label{lagra}
\end{equation}
The second derivative in the exponent comes from the correct discretisation (see box).
We recognise the action of a polymer -- the monomer index
being the time -- at temperature $T$ in a potential $\propto |\nabla
V|^2$. This analogy can be exploited fruitfully~\cite{ChW}.

\vspace{.2cm}

\fbox{\parbox{12cm} {%
{\bf Conventions:}

Suppose we had started from
    (\ref{hermi1}) and constructed the Lagrangian path integral as in
    quantum mechanics textbooks. We would have obtained (\ref{lagra}),
    the  exponentials of the potential at the ends of the trajectory
 being just the change of basis leading
    to (\ref{hermi1}). The  extra term $ - \frac{1}{2} \int \;
    dt \; \frac{\partial^2 V}{\partial q_i^2}          $ in the exponent appears naturally.
    In our case, it appeared as part of the integral $\int dt  \; \dot q_i \left(  \frac{\partial V}{\partial q_i}     \right)$,
   once we specify what this means in terms of a discrete sum.   

The question of discretisation conventions must be treated with care, as they may
lead to errors in the results. However, one should not exaggerate the physical importance
of the whole problem, as it dissappears as soon as the system is regularised by including inertia.  }}

\vspace{.2cm}

\subsection{Kramers equation}

All the steps above can be performed for a system with inertia. A
somewhat confusing fact is that our extended space consists of $4N$
coordinates, including the $2N$ original coordinates {\em and momenta}
$({\bf q},{\bf p})$ playing the role of coordinates of the extended
space, and the  $2N$ 
`hat' variables $({\bf \hat q},{\bf \hat p})$,
 playing the role of momenta in the
extended space,  associated now with the operators:
\begin{equation}
\hat q_i^{op} \rightarrow T \frac{\partial}{\partial q_i} \;\;\;\; ; \;\;\;\;
\hat p_i^{op} \rightarrow T \frac{\partial}{\partial p_i}
\end{equation}

\chapter{Time-reversal and Equilibrium}

As we have seen in the previous section, stochastic dynamics lead to
an evolution that is quite close to a generic quantum problem, often
with a non-Hermitian Hamiltonian. Such generality is in a way bad news, since
one can hardly expect to find a  method to solve a problem that is so general. 
A special and important class of that of systems  that evolve
 in contact with an equilibrium thermal bath, and are under the
action of conservative
forces. There is then a symmetry that can be interpreted  as
time-reversal, which has important consequences.  
We shall also see that if this symmetry is broken by forces that do work, the symmetry-breaking
term can be interpreted as an entropy production rate: this will be the basis of the Nonequilibrium
Theorems of section 6.

\section{Detailed Balance}

The detailed balance property is a relation between the probabilities
of going  from a configuration $a$ to a configuration $b$ and
vice-versa:
\begin{equation}
e^{-\beta V(a)} P_{a \rightarrow b} = e^{-\beta V(b)} P_{b \rightarrow a}
\label{detailed}
\end{equation}
The name `detailed' comes from the fact that if we only ask for the
Gibbs-Boltzmann distribution to be stationary, we need only that
(\ref{detailed}) holds added over configurations, and not term by
term.  We can telescope (\ref{detailed}) to obtain for a chain of
configurations:
\begin{equation}
 P_{a_1 \rightarrow a_2}  P_{a_2 \rightarrow a_3} \dots 
 P_{a_{m-1} \rightarrow a_m}= e^{-\beta [V(a_m)-V(a_1)]} 
P_{a_m \rightarrow a_{(m-1)}}  \dots  P_{a_3 \rightarrow a_2}
  P_{a_2 \rightarrow a_1}  
\end{equation}
which means that 

\vspace{.2cm}
\begin{center}
{\bf Probability [path]}= $\;\;
e^{-\beta[V({\mbox{final}})-V({\mbox{initial}})]}\;\;$ {\bf
  Probability [reversed path]}
\end{center}

\vspace{.2cm}

  And, in particular, for all closed
circuits:
 
\vspace{.2cm}

\begin{center}
{\bf Probability [circuit]} = {\bf Probability [reversed circuit]}
\end{center}

\vspace{.2cm}

In other words, we have the Onsager-Machlup reversibility:
\begin{itemize}
\item The probability of any path going from {\bf $a$} to {\bf $b$} is
  equal to the probability the time-reversed path, times a constant
  that only depends on the endpoints {\bf $a,b$}.
\item Hence, if for some reason there is essentially one type of path
  that takes from {\bf $a$} to {\bf $b$}, then there is also
  essentially one path that takes from {\bf $b$} to {\bf $a$}, and it
  is its time-reversed.
\end{itemize}
These properties are directly observable experimentally, see
\cite{CilibertoGaspard}.

\subsection{Fokker-Planck}

Detailed balance holds for the Fokker Planck evolution without forcing
$f_i=0$. Let us see the implication this has:
\begin{eqnarray}
\langle {\bf q' }|e^{-tH_{FP} } | {\bf q } \rangle \; e^{-\beta V({\bf q })}&=& 
\langle {\bf q  }|e^{-tH_{FP} } | {\bf q' }\rangle  \; e^{-\beta V({\bf q' })}
\nonumber \\
\langle {\bf q' }|e^{-tH_{FP} }  e^{-\beta V}| {\bf q } \rangle &=& 
\langle {\bf q  }|e^{-tH_{FP} } e^{-\beta V} | {\bf q' }\rangle
 = \langle {\bf q'  }|e^{-\beta V} e^{-tH^\dag_{FP} }  | {\bf q }\rangle
\end{eqnarray}
Since this is true 
$\forall ({\bf q},{\bf q'})$, we have:
\begin{equation}
 e^{\beta V} e^{-tH_{FP} }  e^{-\beta V}= e^{-tH^\dag_{FP} } \;\;\; \forall t \;\;\;
\;\;\; \rightarrow \;\;\;\;\;\;\;\;\;
 e^{\beta V} H_{FP}   e^{-\beta V}= H^\dag_{FP}
\label{prehermi} 
\end{equation}
which in turn means that:
\begin{equation}
H_h= e^{\beta V/2} H_{FP}   e^{-\beta V/2}
\label{hermi}
\end{equation}
 is Hermitian, a fact that we have already checked (cfr (\ref{hermi})).
Equation  (\ref{prehermi}) gives also a direct relation between
right and left eigenvectors:
\begin{equation}
|\psi^{L}_\alpha\rangle  = e^{\beta V} |\psi^{R}_\alpha\rangle
\label{since}
\end{equation}
We now understand why an Hermitian form cannot be obtained  when there is forcing:
detailed balance is then lost.

\subsection{Kramers}

Let us now see how this generalises to a process with inertia, having
an energy ${\mathcal H}=\frac{{\bf p}^2}{2m} + V({\bf q})$.  In this
case, something like detailed balance holds, but on the condition that
we reverse the velocities:
\begin{equation}
e^{-\beta {\mathcal H}(a)} P_{a \rightarrow b} = e^{-\beta {\mathcal
    H}(\bar b)} P_{\bar b \rightarrow \bar a}
\label{detailedk}
\end{equation}
where $\bar a$, $\bar b$ are the configurations $a,b$ with the
velocities reversed. Also in the Kramers case we can telescope (\ref{detailedk}) to obtain
relations for trajectories, and for closed circuits, as we did above for the Fokker-Planck
evolution. 

 In operator notation, Eq. (\ref{detailedk})  reads:
\begin{eqnarray}
\langle {\bf q', p' }|e^{-tH_{K} } | {\bf q,p } \rangle \; 
e^{-\beta  {\mathcal H}({\bf q,p })}&=& 
\langle {\bf q, -p }|e^{-tH_{K} } | {\bf q',-p' }\rangle 
\; e^{-\beta  {\mathcal H}({\bf q' })}
\end{eqnarray}
which leads to:
\begin{equation}
 \Pi e^{\beta {\mathcal H}} \; H_{K} \; e^{-\beta {\mathcal H}}
 \Pi^{-1} =\; H^\dag_{K}
\label{hermik}
\end{equation}
where we have introduced the operator that reverses velocities $ \Pi
{\bf p} \Pi^{-1} = -{\bf p} $, and similarly with derivatives.  $H_K$
cannot, in general, be taken to an Hermitian form.  Applying to (\ref{hermik}) steps
analogous to those leading from (\ref{prehermi}) to (\ref{hermi}) one
finds that Hermiticity is broken because $\Pi^{1/2}$ is not real.

\fbox{\parbox{12.5cm} {%

{\bf Another form of time reversal: the adjoint}

A driven overdamped system admits a form of 
time-reversal~\cite{Bertini1,Chetrite}
that is not, however, a symmetry.
Consider the Fokker-Planck operator with nonconservative forces,
and assume we know its stationary distribution:
\begin {equation}
H|stat\rangle =
-\frac{\partial }{\partial q_i} \left(T\frac{\partial}{\partial q_i} +
f_i\right)|stat\rangle =0
\end{equation}
Put $\langle {\bf q}|stat \rangle \equiv \phi({\bf q})$ and compute:
\begin{equation}
\phi^{-1} H \phi =-\left[T
\frac{\partial }{\partial q_i}+ 
\left( 2T \frac{\partial \ln \phi}{\partial q_i}
+ f_i \right) \right]\frac{\partial }{\partial q_i}= H_{adj}^\dag
\end{equation}
Where we have defined the adjoint 
\begin{equation}
H_{adj}= -\frac{\partial }{\partial q_i}\left[T
\frac{\partial }{\partial q_i}-
\left( 2T \frac{\partial \ln \phi}{\partial q_i}
+ f_i \right) \right]
\end{equation}
This  describes another diffusion problem at temperature $T$ 
in a new force field:
\begin{equation}
f_i^{rev}=-\left( 2T \frac{\partial \ln \phi}{\partial q_i}+ f_i \right)
\end{equation}
which only coincides with the original one when ${\bf f}$
derives from a potential.
This formula is the basis of the Hatano-Sasa formula~\cite{Hatanosasa}. 

A  similar but 
stronger form of time-reversibility arises if we accept that
some variables change signs, as velocities do. We refer the reader to
Ref. \cite{Graham}.
Note that in all these cases, we need to know the stationary distribution 
{\em a priori}, so the formulas are moderately useful in practice. 
}}

\section{Equilibrium theorems:  Reciprocity and Fluctuation-Dissipation}

Detailed balance is a form of time reversal symmetry in the trajectories.
 It cannot come as a surprise that in equilibrium it implies time-reversal
symmetry in the correlation functions. Let us do it for the Kramers equation.
Denoting $|GB\rangle$ the Gibbs-Boltzmann distribution: 
\begin{eqnarray}
C_{AB}(t-t') &=& \langle - | A \;e^{-(t-t')H_K} \; B  | {{GB}} \rangle=
\langle GB |  B \; e^{-(t-t')H^\dag_K} \; A | - \rangle \nonumber \\
&=&\langle - |  e^{-\beta {\mathcal H}} B  
\; e^{-(t-t')H^\dag_K} A \; e^{\beta {\mathcal H}} | GB \rangle \nonumber \\
&=&
\langle - | \underbrace{\Pi e^{-\beta {\mathcal H}} B 
 e^{\beta {\mathcal H}} \Pi^{-1}}_{\bar B} \;e^{-(t-t')H_K} \; 
 \underbrace{\Pi e^{-\beta {\mathcal H} } 
 A  e^{\beta {\mathcal H}} \Pi^{-1}}_{\bar A}  
 | GB \rangle \nonumber \\
&=&C_{\bar B \bar A} (t-t')
\end{eqnarray}
where we have used (\ref{hermik}) and we have defined $\bar A({\bf q},{\bf p})=
 A({\bf q},-{\bf p})$.
The same derivation can be done for the 
Fokker-Planck case for observables dependent on coordinates only.

Another important equilibrium property is the  Fluctuation-Dissipation theorem.
It relates the response of the expectation
of an observable $A$ produced by a kick given by 
field conjugate to an observable $B$ to the corresponding two-time correlation.
Putting ${\mathcal H}_{h_B} = {\mathcal H} - h_B B $, we compute:
\begin{equation}
R_{AB}(t-t') = \frac{\delta \langle A(t)\rangle }{\delta h_B(t')}=
\langle - | A \;e^{-(t-t')H} \; \left.\frac{dH}{dh_B}
\right|_{h_B=o} | 
{\mbox{GB}} \rangle
\label{popopo}
\end{equation}
Because the equilibrium distribution in the presence of a field 
 $ |{GB} \;(h_B) \rangle$ satisfies for all $h_B$ that 
$H_{h_B} |{GB} \;(h_B) \rangle =0$ we have:
\begin{eqnarray}
\left. \left(\frac{dH} {dh_B}\right) | {{GB}} \rangle \right|_{h_B=0} &=& 
- H \left. \left(\frac{d}{dh_B}   | {GB}   \rangle\right) \right|_{h_B=0} 
=
\beta  H  \; (B - \langle B \rangle) \;
 | {GB}   \rangle_{h_B=0}\nonumber \\
&=& \beta \; HB  \; | {GB}   \rangle_{h_B=0}
\label{iii}
\end{eqnarray}
and substituting in (\ref{popopo}):
\begin{eqnarray}
R_{AB}(t-t') 
&=& \beta \langle - | A \;e^{-(t-t')H} \; 
 H  \; B \;|
{{GB}} \rangle \nonumber \\
&=& \beta \frac{\partial }{\partial t'} C_{AB}(t-t')
\label{FDT}
\end{eqnarray}

\subsection{FDT of the first and second kind. A derivation of
the Langevin equation.}

At the outset we started with a bath that contained friction proportional to
$\gamma$ and noise whose variance is $\gamma T$. This very precise relation
between noise and friction
is often refereed to as `Fluctuation Dissipation of the first kind',
because it relates the dissipation and fluctuations {\em of the bath},
rather than of the system.
If the bath satisfies this relation, a system in contact with it will 
eventually equilibrate (this might take long) and will then
verify the fluctuation dissipation theorem (\ref{FDT}) -- of the `second kind' 
-- for its observables.
 
Next, assume there is a large number $\alpha=1,...,M$ of independent copies of such systems,
with coordinates ${\bf q}^\alpha,{\bf p}^\alpha$, all in equilibrium
with a bath. We  intend to use them in turn  as baths for a 
further system of coordinates    ${\bf q}',{\bf p}'$ and energy ${\mathcal
H}'({\bf q}',{\bf p}')$. To do this we couple them, for example through a term:
\begin{equation}
{\mathcal H} = \underbrace{\sum_\alpha {\mathcal H}_\alpha}_{bath} +  \underbrace{
{\mathcal H}'({\bf q}',{\bf p}')}_{system} - 
\underbrace{M^{-\frac{1}{2}} \sum_\alpha {\bf q}^\alpha 
{\bf .} {\bf q}'}_{coupling}
\end{equation}
 We may ask what is the condition for the coupling term to constitute a 
legitimate thermal bath for the primed system. 
The equations of motion of the primed variables are: 
\begin{equation}
\ddot q^{'} = -\frac{\partial \mathcal H'}{\partial q_i^{'}} - {\bf h(t)}
\end{equation}
Where the field ${\bf h}$ is 
$ {\bf h}=M^{-\frac{1}{2}}\sum_\alpha q_\alpha$.
The large $M$ limit allows us to treat each $M^{-\frac{1}{2}}  {\bf q}^\alpha 
{\bf .} {\bf q}' $ as a small perturbation to the system ${\mathcal H}_\alpha$, and to invoke
the central limit theorem to say that  $M^{-\frac{1}{2}}  {\bf q}^\alpha$ is a Gaussian.
 Assuming the expectation values of 
$\langle q_\alpha \rangle=0$ in the absence of coupling, the field ${\bf h}$ has two contributions
for large $M$: {\em i)} a random Gaussian noise $\eta(t)$ with correlation $C_{\alpha \alpha}(t,t')=
\langle \eta(t) \eta(t') \rangle = \langle q_\alpha(t) q_\alpha(t')\rangle$ and {\em ii)} a drift
due to the back effect of the $q'$ which acts as a field  on the $q_\alpha$. Again, because
 $M$  is large, the average response of the ensemble $\alpha$ is: 
\begin{equation}
M^{-\frac{1}{2}} \langle q^\alpha \rangle = \int_0^t \;  dt' \; R_{\alpha \alpha}(t,t') q'(t')
\end{equation}
The equation of motion of the primed variable becomes:  
\begin{equation}
\ddot q^{'} = -\frac{\partial \mathcal H'}{\partial q^{'}} +\eta(t) + 
\int_0^t \;  dt' \; R_{\alpha \alpha}(t,t') q'(t')
\end{equation}
This is in fact the generalised Langevin equation, describing the
primed system in contact with a thermal bath with coloured noise and
friction with memory~\footnote{The white noise limit  (\ref{lange1})  is obtained when
$C_{\alpha\alpha}(t-t') \propto \delta(t-t')$.}.  The condition that
the bath is a good equilibrium one is precisely:
\begin{equation} 
T R_{\alpha \alpha}((t,t') = \frac{\partial}{\partial t'} C_{\alpha \alpha}(t,t')
\end{equation}
The primed system will, under the action of this dynamics, equilibrate to
the Gibbs-Boltzmann distribution~\footnote{With an energy that includes a contribution $\langle [q^{\alpha}]^2  
\rangle q^{'2}$ coming
from the interaction term (The factor $\langle [q^{\alpha}]^2 \rangle $ is the
equilibrium expectation  for a single isolated ${\mathcal H}_\alpha$).
 This term can compensated by an opposite one in ${\mathcal H}'$}.
\begin{itemize}
\item
 Fluctuation dissipation of the second kind for the $q_\alpha$ has become a fluctuation dissipation
of the first kind when they are considered as a bath for the primed variable $q'$.
\item
Friction is the back reaction of the bath to the perturbation exerted by the system,
while noise is given by the incoherent addition of motions of the bath's constituents.
\item
The fluctuation-dissipation relation regulates the balance between these two effects in such a
way as to guarantee that equilibrium is `passed on' to a coupled system.
\end{itemize}

\vspace{.2cm}

\section{Time-reversal violations and Entropy production}

In the previous section we have seen that there is a symmetry 
related to time-reversal in systems that are in contact with 
an equilibrium thermal bath, and have only conservative forces.
Once forcing is allowed, this symmetry is broken:  in an 
interesting manner.  As we shall now see, the symmetry-breaking term turns out
to  be proportional to the power injected by the nonconservative forces,
divided by a temperature: it can be interpreted as an entropy 
production~\footnote{Subtle questions about when can one call this entropy are
extensively addressed in the literature~\cite{Maes_entropy}, here we shall not
get into these matters}.  

For simplicity, we shall do this calculation in two cases with
inertia, in order to avoid unnecessary
complications brought about by the fact mentioned above that
 the power in an overdamped case is a
discontinuous function of time.

\vspace{.1cm}

{\bf Kramers equation}

\vspace{.1cm}

Let us attempt to obtain a relation like (\ref{hermik}) in the presence of
forcing. Referring to (\ref{Hk}) $H_K=H_{Liouville}+H_b+H_f$ we compute
 $\Pi  e^{\beta E} H_{K}   e^{-\beta E} \Pi^{-1} $
in detail. First, we make the similarity transformation:
\begin{eqnarray}
   e^{\beta {\mathcal H}} H_{Liouville} e^{-\beta {\mathcal H}} &=&
   H_{Liouville} \nonumber \\ e^{\beta {\mathcal H}} H_{b}
   e^{-\beta {\mathcal H}} &=& \gamma \left(T\frac{\partial}{\partial
     p_i}-\frac{p_i}{m} \right)\frac{\partial}{\partial p_i} \nonumber
   \\ e^{\beta {\mathcal H}} H_{f} e^{-\beta {\mathcal H}} &=& - f_i
   \left(\frac{\partial}{\partial p_i}-\beta \frac{p_i}{m} \right)
\end{eqnarray}
Next, Hermitian conjugation changes signs of derivatives, and reverses
the order of factors. Finally, velocity reversal transforms $(p_i
\rightarrow -p_i)$ and $\frac{\partial }{\partial p_i} \rightarrow
-\frac{\partial }{\partial p_i} $.  All in all we get:
\begin{equation}
\left[ \Pi e^{\beta {\mathbf H}} H_{K} e^{-\beta {\mathbf H}} \Pi^{-1}
  \right] =\; H^\dag_{K} - \underbrace{{\beta} \sum_i f_i
  v_i}_{\frac{POWER}{T}}
\label{wok}
 \end{equation}
(here again, $v_i=p_i/m$ is the velocity).
As announced, we find that relation  (\ref{hermik}) has now an extra term
corresponding to power divided by temperature, an entropy production rate.

\vspace{.1cm}

{\bf Gaussian Thermostat}

\vspace{.1cm}

We may repeat the calculation for the Gaussian thermostat (\ref{eq3}).
Only velocity reversal is necessary, and we obtain:
\begin{eqnarray}
\left[ \Pi H_G \Pi^{-1}\right]^\dag &=& H_G +  [\gamma p_i]\frac{\partial}{\partial p_i}
-\frac{\partial}{\partial p_i}\; [\gamma p_i] \nonumber\\
  &=&  H_G   -(N-1)
 \frac{{\bf f}\bullet {\bf p}}{{\bf p^2}}
\label{dodo}
\end{eqnarray}
(summation convention) 
where $2N$ is the dimension of phase-space. Again,
this has the interpretation of a power divided by a kinetic temperature $(\sim 
{\bf p}^2)$.

\section{What does a stationary out of equilibrium distribution
look like?} \label{pp}

A driven system, if coupled to a thermostat will reach a stationary
distribution.  The problem with out of equilibrium statistical
mechanics is that there is no simple, general expression for this distribution.
 If the thermostat keeps the energy constant, as the
Gaussian one we defined above, even if the probability 
is then restricted to an energy shell, it does not
cover it  in a uniform, microcanonical way.  In fact, if the
thermostat is deterministic, the distribution will be a fractal.
If, on the other hand, the thermostat involves noise, the fractal will
be blurred, but the distribution is still non-uniform on the energy shell.
Can we get some intuition on this? The
purpose of this section is to see, in a relatively simple
example~\cite{Bonetto,Kur}, what happens  when a system
 is forced.

\begin{figure}[htbp]
\begin{center}
\includegraphics[width=7cm]{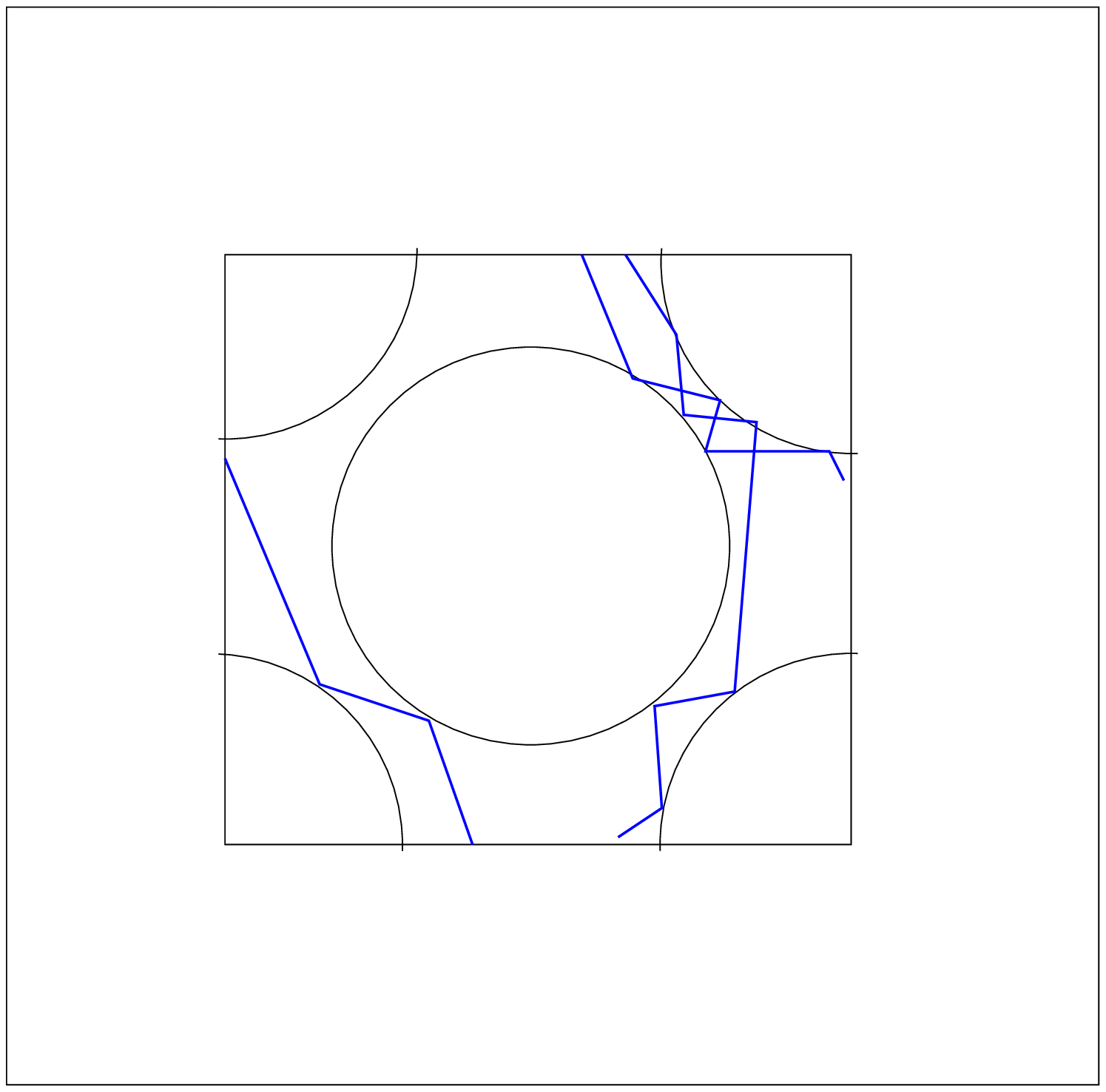} \hspace{.1cm}
\includegraphics[width=7cm]{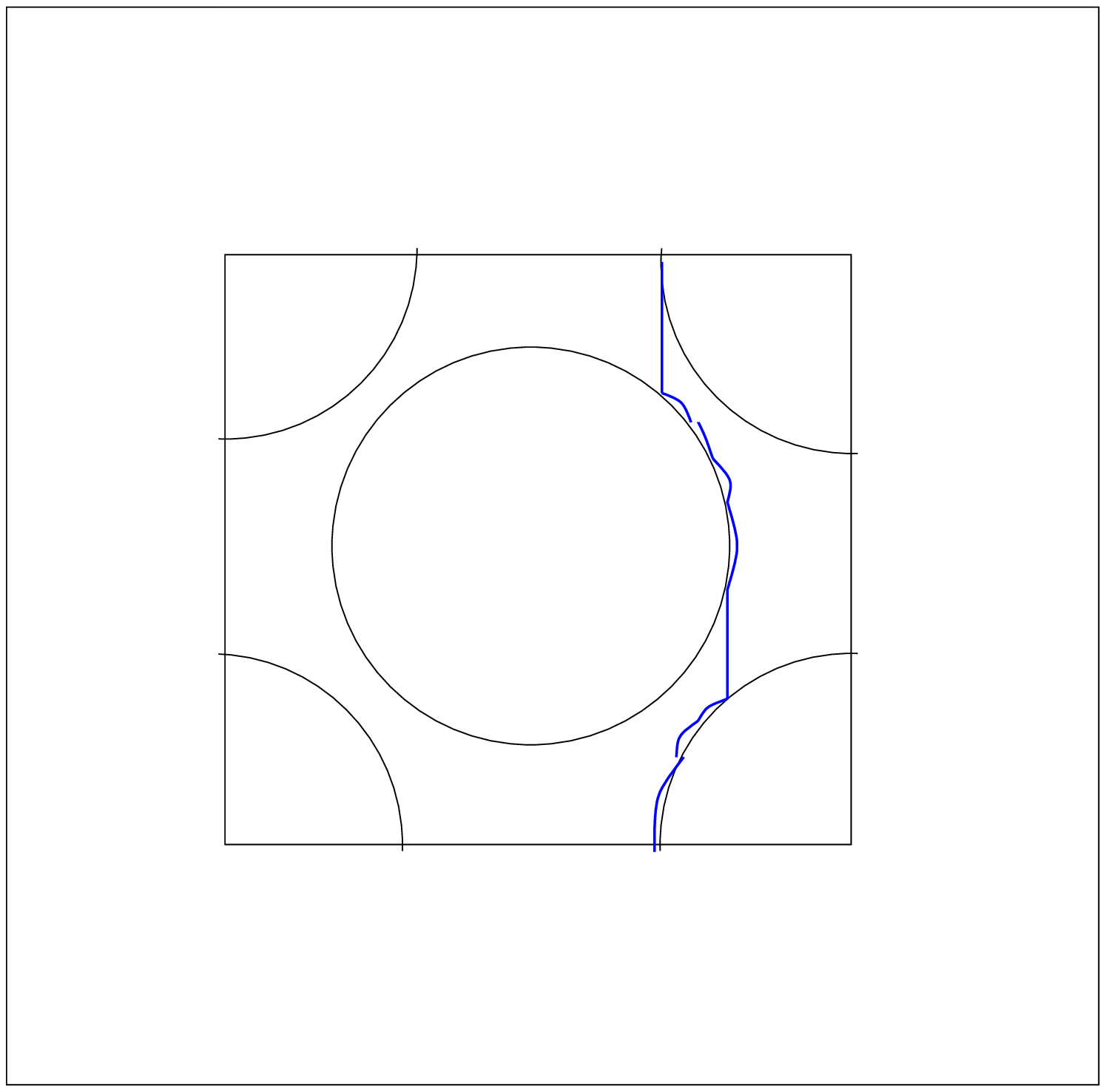}
\end{center}
\caption{Left: a trajectory at zero field. Right: trajectory under a
  very strong field, pointing in the downward direction: the particle follows, through short bounces, the
  surface of the obstacle, and escapes when the field becomes tangent
  to it.}
\label{billiard}
\end{figure}

\begin{figure}[htbp]
\begin{center}
\includegraphics[width=12cm]{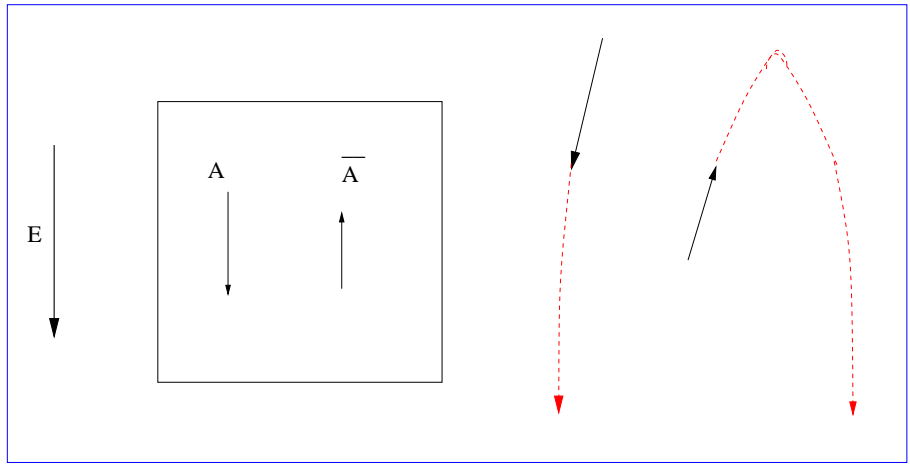}
\end{center}
\caption{Free trajectory under a field. Attractor $A$ and repellor $\bar A$ are parallel and
antiparallel to the field.}
\label{Free}
\end{figure}

We shall consider a Lorentz gas, or, equivalently, a particle in a
billiard as in Figure \ref{billiard}, under the effect of a constant
field, with periodic boundary conditions.  We  assume that there
is a Gaussian thermostat, which fixes the velocity modulus to be
constant, that we can take as unity.
The trajectories between bounces are given by:
\begin{eqnarray}
\dot p_x &=& E - \frac{E p_x^2}{p^2} 
\nonumber \\
\dot p_y &=&  - \frac{E p_x p_y}{p^2}\nonumber \\
 p \dot \theta &=& -E \sin \theta  = -\frac{d}{d\theta} [-E \cos \theta]
\label{gogo}
\end{eqnarray}
where we have defined the angle $(p_x,p_y)= p(\sin \theta,\cos \theta)$ (Figure \ref{Free})
If there were no obstacles, there would be two stationary situations: when the
velocity is parallel to the field (stable) and when it is antiparallel
(unstable).  Such trajectories constitute the {\em attractor} and the
{\em repellor} respectively.

Consider now the effect of obstacles.  When the field is off, in the
presence of obstacles the system is known~\cite{Sinai} 
to be ergodic.  Phase-space
points on the energy shell -- the Cartesian product of the allowed
configurations times the velocity sphere, are visited uniformly.  On
the other extreme, if the field is very strong, the trajectories stay
bouncing close to the surface of 
 the obstacles until they escape along a tangent direction
parallel to the field, only to hit a new obstacle -- see the right
panel of Figure \ref{Free}.  If we consider a stationary situation, 
these `trickling down'
  trajectories involve {\em only a very restricted part of
  configuration space}.
Adding energy-conserving noise to (\ref{gogo}) does  not change 
dramatically the situation.
In conclusion, the stronger the forcing field $E$, the more focussed 
the {\em attractor}
 reached at long times is on a subset of the energy shell.
As we shall see later, the Gallavotti-Cohen fluctuation relation is,
in a certain sense, a measure of this focussing. 

\section{Spectra of Fokker-Planck, Kramers and Liouville operators}

As we have seen above, the detailed balance property implies that the
Fokker-Planck operator can be taken to an Hermitian form $H_h$ via the
similarity transformation (\ref{hermi}). This implies that its eigenvalues are
real.
If, on the other hand, detailed balance is violated, then there is no
such transformation, and some eigenvalues come in  complex conjugate pairs.
In any case, the Perron-Frobenius theorem mentioned above
implies that the real parts of eigenvalues are non-negative.

On the other extreme, the Liouville operator corresponding to pure
Hamiltonian dynamics
\begin{equation}
H_{Liouville}=\frac{\partial \mathcal H}{\partial p_i} \frac{\partial
}{\partial q_i} - \frac{\partial \mathcal H}{\partial q_i} \frac{\partial
}{\partial p_i}
\end{equation}
(summation convention)
would seem to be, at least superficially, {\em anti}-Hermitian, since it has only
first derivatives.  Its spectrum would be pure imaginary, except that
we have to be careful when we define the space of wavefuncions on which
it acts.
Clearly, the Kramers operator, which in a sense interpolates between
both, is neither Hermitian nor anti-Hermitian, and has pairs of
complex eigenvalues even in the conservative case, as we have mentioned already.

  Consider a strongly chaotic, 
`mixing' Hamiltonian system. If we start from
an  initial configuration  distributed in a small
probability cloud, the probability distribution mixes completely
in phase-space -- like a drop of ink in a stirred liquid, hence the name.
Another way of obtaining the same result is to start from a single
 configuration, but subject the dynamics to a small noise.
The fast mixing of the probability implies a concomitant fast 
decay of correlation functions to their stationary value. As discussed above
(see Eq. (\ref{gap0})), this would imply that there
is a gap in the real part of the spectrum.
 Hence, if the deterministic system is sufficiently chaotic,
we expect that, in the presence of a  small amount of noise,
its spectrum will have a gap in the real part of the lowest eigenvalues. 
When the noise is strictly zero the situation is more subtle, as there are
many unstable periodic orbits, on which the correlations are of course periodic in time.

A very interesting question is then what happens  when we start from a
 Hamiltonian system with stochastic noise, and gradually 
decrease the noise's intensity. 
This can be done with the Kramers equation without
forcing, or, better, with the energy-conserving thermostat
(\ref{eq3}), letting $\epsilon \rightarrow 0$.  The remarkable result,
consistent with the discussion above, 
is that, if the system is sufficiently chaotic, as we let $\epsilon
\rightarrow 0$ some  eigenvalues do not become imaginary: they retain a
positive real part.  These eigenvalues, and the corresponding
eigenvectors, are the Ruelle-Pollicott resonances~\cite{RuellePoll}:
 the ones that have
smaller real part are responsible for the long time  relaxation to
equilibrium.
A simple example of this phenomenon  is discussed in the box below.

\vspace{.2cm}

\fbox{\parbox{12cm} {%

    An instructive example is the particle in a harmonic potential with noise.
The diagonalisation of the corresponding Kramers operator
is simple, and can be found in Risken's book
\cite{Risken}, Chapter 10: The  spectrum is as
follows. The eigenvalues are labelled by $n_1,n_2=0,1,2,...$:

$\bullet$  {\bf Stable case $V=\frac{ 1}{2} \omega^2 q^2$}: 

Eigenvalues $\lambda_{n_1,n_2} = \frac{1}{2} \gamma (n_1+n_2) + \frac{i}{2}
\; \sqrt{4 \omega^2 - \gamma^2} \; (n_1-n_2)$
 The spectrum becomes imaginary
in the $\gamma \to 0$ limit of zero coupling to the bath, when the system becomes an undamped
harmonic oscillator. This is compatible with the existence of many stable, periodic orbits.

$\bullet$  {\bf Unstable case $V=-\frac{ 1}{2} \omega^2 q^2$}: 

Eigenvalues $\lambda_{n_1,n_2} = \frac{1}{2} \sqrt{\gamma^2+4
  \omega^2} \; (n_1+n_2+1) + \frac{1}{2} \gamma (n_1-n_2-1)$
All eigenvalues are real and larger than zero. The latter is to be
expected, given that there is no stationary state. The fact that even
in the limit $\gamma \to 0$ the spectrum stays real may come as a
surprise: it underlines the fact that although the Liouville operator
seems superficially anti-Hermitian (and would thus lead us to expect
pure imaginary eigenvalues), in fact the Hilbert space in which it
acts makes it not be so. Again, we find that the spectrum retains a
real part as $\gamma \to 0$ when it is has {\bf unstable} orbits
($q(t)=0$, $p(t)=0$, in this case) that are destroyed by the any amount of
 noise ~ (see also Refs. \cite{gaspard_nicolis,gaspard_pollicott}).
}}

\chapter{Separation of timescales}

Many systems have processes happening in very different timescales.
Often, what is interesting is what takes long to happen,
while the rapid fluctuations are relatively featureless.
Consider the following examples:

\begin{itemize}
\item {\bf Metastability:} 
Chemical reactions have often metastable states. 
Consider a mixture of Oxygen and Hydrogen. It takes
a very short time for this gas to become `equilibrated' into a mixture
$O_2 +  2H_2$,  staying in
a stationary state until the reaction $O_2 +  2H_2 \rightarrow 2H_2O$
starts somewhere -- an extremely unlikely event --
 and then rapidly propagates throughout. The true equilibrium
state, water vapour, is then reached. 

Similarly, diamond eventually decays
into graphite, but this process is fortunately slow.
\item {\bf Hydrodynamic limit:} Systems with soft modes have  slow evolution
along these modes, while the `hard' ones relax much faster.
The typical example is a liquid, whose macroscopic motion is visible
and slow, while the  density fluctuations at the molecular scale evolve fast.
As we shall see below, in some cases one can make a  
`hydrodynamic' description, with the fast fluctuations acting as a noise
whose intensity goes to zero with the coarse-graining scale.
\item{\bf Coarsening and glasses:} Suppose one quenches a ferromagnet
to a low, but non-zero temperature. Domains of positive and negative 
magnetisation start growing.  Inside each domain, the system resembles a pure
ferromagnetic state with fast magnetisation fluctuations. The domain walls,
 however, evolve slowly, the slower the larger the domains.
A more subtle case is the one of glasses, which have a fast evolution 
(`cage motion') and  slow, collective rearrangements (`aging').
\end{itemize}

In all these cases, our ambition is to concentrate as much as possible
in what is slow and interesting. In some cases, when we know {\em a priori}
which are the slow coordinates, we may attempt to eliminate
 the fast fluctuations by allowing them to thermalise, at fixed value 
of  the slow coordinates which we then treat adiabatically.
One thus obtains a `free-energy landscape' for the slow variables (see e.g. ~\cite{Eijnden}). 

What happens when we do not know exactly who is fast and who is slow?
In the  next sections we introduce a general  approach to metastability,
first in detail in a simple warming-up context, and then mention briefly
how it works in general. In the last section,
we  describe the hydrodynamic limit of a transport problem,
and how it takes us to a low-noise (quasi-deterministic) situation. 

\section{Metastability}

\subsection{The simple case of weak noise}

Consider an overdamped  Langevin system. We wish to analyse
the spectrum of its Fokker-Planck evolution operator in
the weak-noise limit. Because we know that $T$ plays a role analogous
to $\hbar$, we shall use what we know from semiclassical Quantum Mechanics.
Let us first transform $H_{FP}$   to its Hermitian basis (\ref{hermi}):
\begin{equation}
H_h= e^{\beta V/2} \; H_{FP} \; e^{-\beta V/2} = \frac{2}{T} \sum_i
 \left[ - \frac{T^2}{2} \frac{\partial^2}{\partial q_i^2} +
 \underbrace{ \frac{1}{8} \left(\frac{\partial V}{\partial
 q_i}\right)^2 - \frac{T}{4} \frac{\partial^2 V}{\partial
 q_i^2}}_{V_{eff}} \right]
\label{hermi11}
\end{equation}
Consider first  a one-dimensional  harmonic potential:
\begin{equation}
V=\frac{1}{2} a q^2  \;\;\; ; \;\;\;  H_{FP} = -\frac{d}{dq}\left[ T\frac{d}{dq}
+a q \right]
\end{equation}
We have
\begin{equation}
H_h= e^{\beta V/2} \; H_{FP} \; e^{-\beta V/2} = \frac{2}{T} 
 \left[ - \frac{T^2}{2} \frac{d^2}{\partial q^2} +
  \frac{1}{2} \left(\frac{a}{2}\right)^2 q^2 - \frac{T}{4}  a \right]
\label{hermi2}
\end{equation}
Apart from the global factor $\frac{2}{T}$ we recognise the Hamiltonian
of a Quantum
oscillator with $\omega=|a|/2$ , $\hbar=T$ and $\hbar \omega = Ta/2$. 
The eigenvalues are then $\hbar \omega (n+\frac{1}{2})$, that is:
\begin{equation}
\lambda \;\; \;\; = \frac{2} {T} \left[ \left(n+\frac{1}{2}\right)
\frac{T|a|}{2} - \frac{Ta}{4} \right] = \left\{
\begin{array}{ccc} 
 & 0,\, |a|,\, 2|a| , \,... \;\;\; &if\, a>0 \\
 & |a|,\, 2|a|\, ,\, ... \;\;\;    &if \, a<0 \\
\end{array}
\right.
\label{eige}
\end{equation}
As expected, the lowest eigenvalue is zero in the stable,
 and positive in the unstable case. The gap between eigenvalues
is proportional to the curvature of the potential.

We can extend this result to the stationary point of any potential, using the fact that at low temperature
only the neighborhood of the saddle points contribute. 
Developing $V_{eff}$ around a minimum, which we assume is in $q=0$,
and putting $\frac{q}{\sqrt{T}} \rightarrow x $ we have
\begin{eqnarray}
V_{eff} &=& V_{eff}^{''}(0) \frac{q^2}{2}
\; + \; V_{eff}^{'''}(0) \frac{q^3}{6} + ...\nonumber \\
&=& V_{eff}^{''}(0)
\frac{x^2}{2} \; \underbrace{ + \; V_{eff}^{'''}(0) \sqrt{T}
\frac{x^3}{6}}_{subdominant} \; + ...
\end{eqnarray}
So that
\begin{equation}
H_h = 2 \left[ -\frac{1}{2} \frac{d^2}{dx^2} \; + \; 
 \frac{1}{2} \left(\frac{|V''(0)|}{2}\right)^2 x^2 - \frac{1}{4}  V''(0)
  \;  + \; subdominant \;... \right]
\end{equation}
The eigenvalues are given by (\ref{eige}) with $V''(0)$ playing the
role of $a$.

The generalisation to many dimensions is straightforward. Since one
has to develop the potential only to second order around a saddle,
one can then go to the basis where the second derivative matrix is diagonal,
and treat each mode as an independent oscillator.
Every saddle point yields to this order an independent spectrum in the
Fokker-Planck operator, but only local minima have zero eigenvalues.
There is then exactly 
one zero-eigenvalue per minimum.  If the calculation is
done exactly, one finds that the degeneracy is
 lifted by a small  amount inversely proportional
to the  passage times between states, in this case exponentially small
in the inverse temperature. If there are $p$ local minima, there are then 
$p$  `almost zero' eigenvalues $\lambda_1,...,\lambda_p$ with the
associated escape times $\lambda_1^{-1},...,\lambda_p^{-1}$ bounded by
the smallest escape time $t_{pass} \;=\; \mbox{min} \{\lambda_1^{-1},...,
\lambda_p^{-1}\}$.

\begin{figure}[htbp]
\begin{center}
\includegraphics[width=12cm]{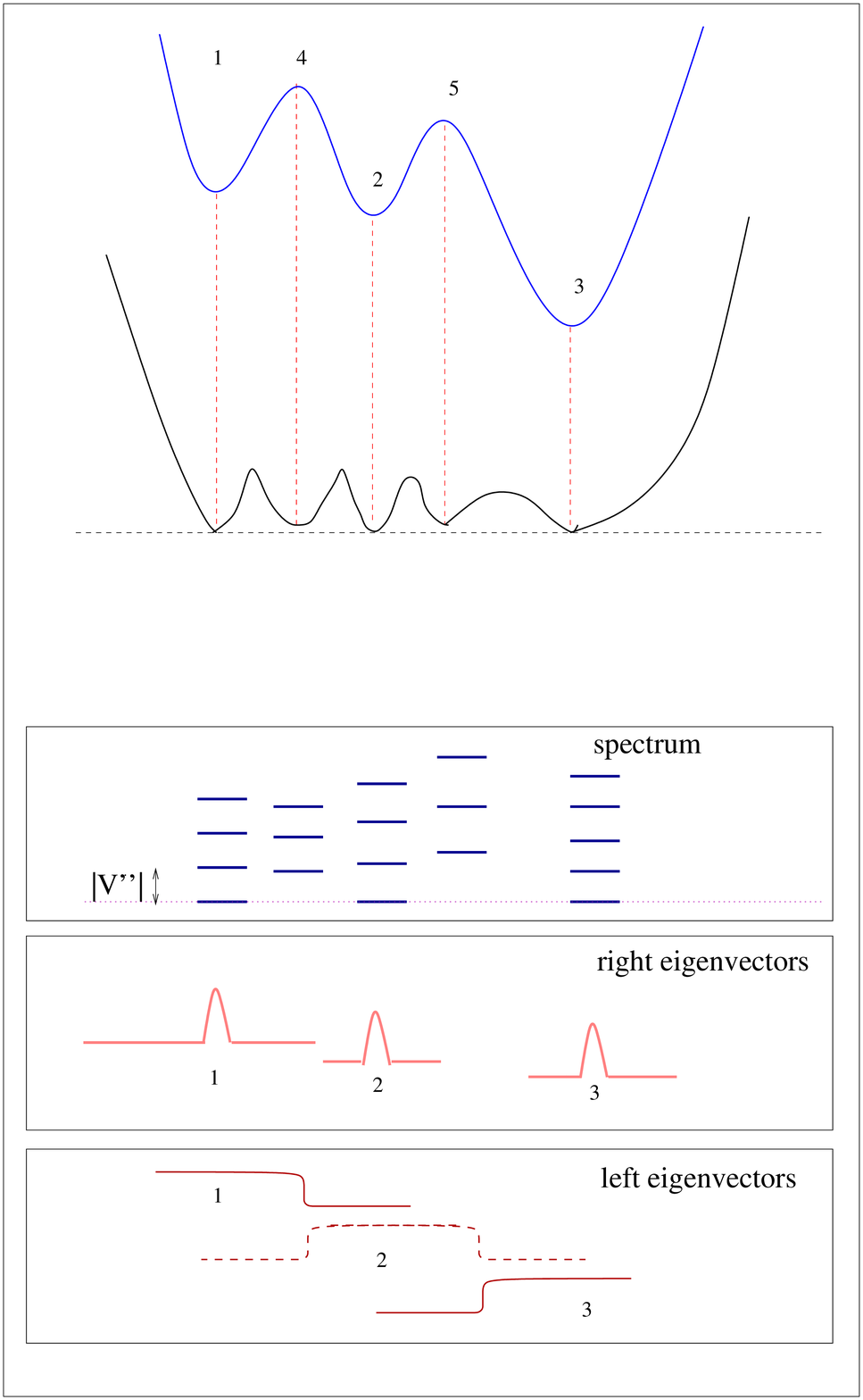}
\end{center}
\caption{ Low-temperature spectrum for a multi-valleyed potential
  $V$(top), and the associated  effective potential $V_{eff}$ (below). 
Only the minima of $V$ contribute with
  near-zero eigenvalues. The near-zero (in this case three-dimensional) subspace is
spanned by right eigenfunctions that  are essentially the equilibrium
  distribution in each basin, and left eigenfunctions (the {\em committors}) that are essentially constant within each
  basin. The places where committors take the value one half
  are the {\em transition states}.}
\label{lab}
\end{figure}

Let us now turn to the eigenvectors corresponding to the 
`almost zero' eigenvalues.  The construction we have done above for
 eigenvalues can be completed to obtain (approximate) right and left 
eigenvectors. Close to a minimum, we expect the right eigenvector to be
correspond to a Gibbs distribution peaked around it. Similarly, we expect the
 left eigenvector to be essentially a constant. This is in fact the case,
but the result is stronger, as it holds throughout the basin
of attraction of each minimum. The situation is depicted in Fig. \ref{lab}.
 For small temperatures, 
it is as if   an infinitely high, thin wall would enclose each basin, within which 
the eigenvectors and eigenvalues are those of the isolated region.
Eventually, the
  finite-temperature corrections which split the
degeneracy of `almost zero' eigenvalues, also mix the approximate eigenvectors
 localised in each basin.  

\subsection{ General approach to metastability. 
Spectra, states and committor functions}

\begin{figure}[htbp]
\begin{center}
\includegraphics[width=12cm]{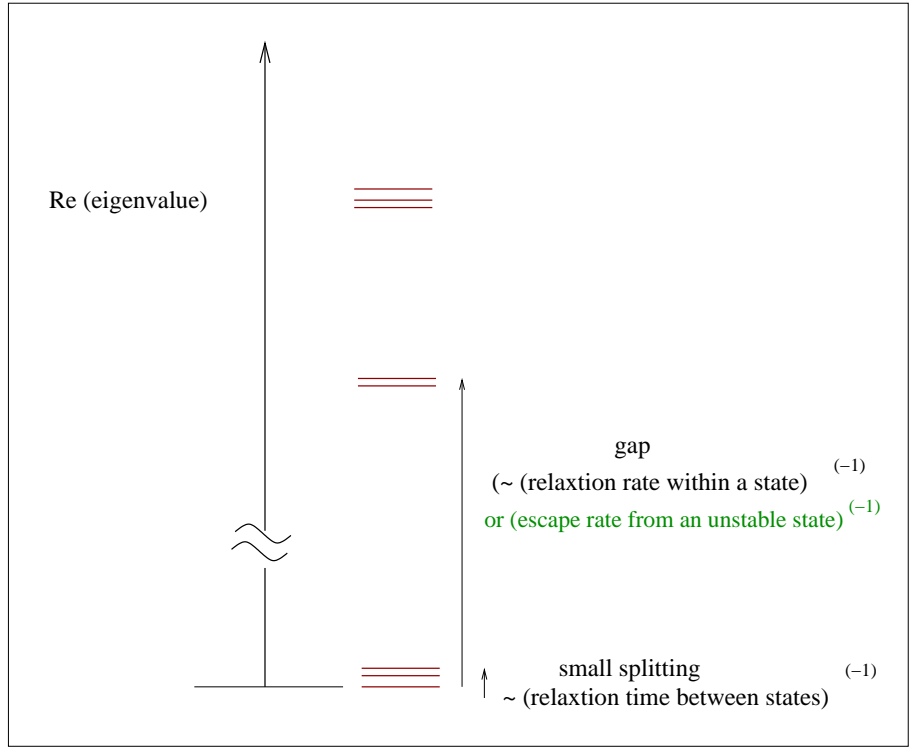}
\end{center}
\caption{The general situation in a system with metastability, 
whatever the origin of timescale separation. The states $|P_a\rangle$ that
can be interpreted as metastable probability distributions are linear
 combinations of the right eigenvectors `below the gap'.}
\label{lab1}
\end{figure}

It turns out that the low-temperature situation is just an instance of
a very general scenario. Indeed , one can turn things around and use
the existence of a gap in the spectrum to give a general and useful
definition of metastability \cite{larry,bovier} (see \cite{biku}, for an
application to metastability in glasses).  Consider
an evolution operator having the lowest $p$ eigenvalues
$\lambda_1,...,\lambda_p$ whose real part is separated by a gap from
the others   $\lambda_{p+1},\lambda_{p+2},...$, (see Fig \ref{lab1}).
There are two characteristic times $t_{pass}= \mbox{ min } \{
\lambda_1^{-1},...,\lambda_p^{-1}\}$ and  $t^*= \mbox{ max } \{
\lambda_{p+1}^{-1},\lambda_{p+2}^{-1},... \}$: they will be interpreted as
the minimal time  needed to escape a metastable state,
 and to equilibrate within
a state, respectively.
In the previous section   $t^*$ is of order one and $t_{pass}$ is exponentially
large in $T$.
At times of order $t\gg t^*$, the dynamics projects completely to the space
below the gap, but if $t \ll t_{pass}$ 
there is still no time for discerning between different
eigenvectors below the gap.
Clearly, the operator $\exp[-tH]$ for $t^*\ll t\ll t_{pass}$  
is essentially a projector onto
the space `below the gap' (up to terms of order $\exp[-t\lambda_a]$,
with $a> p$). 

Within the same accuracy, it turns out that
one can then find a basis of $p$ right 
eigenvectors $|P_a\rangle$ which are:
\begin{itemize}
\item positive:
  $P_a({\bf q}) = \langle {\bf q} |P_a\rangle\geq 0$:
\item almost stationary: $H |P_a\rangle \sim 0 \;\;\;\;\;\; \forall \;
  a=1,...,p $
\item normalised and not zero in non-overlapping regions of space.
\end{itemize}
As a consequence the $|P_a\rangle$ vectors have all the good properties
 of  metastable states: they are positive normalised distributions, 
 non zero only on different regions of the configuration space and 
 are stationary on time scales less than $t_{pass}$. 
The last property is related to the fact that one can
also find a basis of $p$  almost-stationary
($\langle  Q_a| H \sim 0$) left eigenvectors $ \langle  Q_a|$.
They  satisfy the approximate orthogonality and
normalisation conditions:
\begin{equation}  
 \langle  Q_a|P_b\rangle \sim \delta_{ab} 
\label{a}
\end{equation}
One can also write  approximately:
\begin{equation}  
 e^{-t H} \sim \sum_a |P_a\rangle\langle  Q_a|
\label{generator} 
\end{equation} 
Note that neither $ \langle  Q_a|$ nor $|P_b\rangle$ are
exact eigenvectors `below the gap', but linear combinations of them.


The $Q_a({\bf q})$ are the {\em committor functions}~\cite{Dellago,committor}
 of the states,
giving the probability that starting on a certain point ${\bf q}$
the dynamics (again, in times $t^*\ll t\ll t_{pass}$) goes to the state $a$.
This can be easily seen as follows: the probability of ending in a
point ${\bf q}'$ starting from a point ${\bf q}$ is, at times of order
$t^*\ll t\ll t_{pass}$
\begin{equation}
 Probability \; \sim \; \langle {\bf q}'| e^{-tH} |{\bf q}\rangle
 \sim \sum_i \langle {\bf q}'|P_a\rangle \langle Q_a |{\bf q}\rangle
\end{equation}
If the point ${\bf q}'$ is well within the state `$a$', then $P_a({\bf
  q}')$ is large and the other $P_b$ $(b \neq a)$ are small. There is
only one non-zero term in this sum, and we conclude that the
probability to fall in the state `$a$' is proportional to $Q_a({\bf
  q})$.  Each $Q_a({\bf q})=\langle Q_a|{\bf q}\rangle$ is essentially
one within the basin of attraction of the state $a$ , and almost zero
everywhere else.  The places where the $Q_a({\bf q})\sim \frac{1}{2}$
are called the {\em transition states}.

Given  any observable $A$, we can calculate its average within the
state `$a$' as:
\begin{equation}  
\langle A \rangle_a = \langle  Q_a|A|P_a\rangle 
\label{expectation}
\end{equation}

Again,  the situation described above can be summarised by saying that
for times  $t^*\ll t \ll t_{pass}$ 
everything happens as if there is an infinite
wall  enclosing
each basin of attraction.
In the proof, as in the simple example of the previous subsection,
 the definition
is unavoidably linked to the timescales $(t^*,t_{pass})$:
 if one considers really
infinite times, before any other limit, then the distinction between
states vanishes.
In real life, this separation of timescales might be controlled by a parameter,
or it might be just a more or less valid approximation. 
As an example of the former,
consider an Ising ferromagnet of size $L$:
 the longest relaxation within a state is
the time for a domain to grow to the size of the system $t^*\sim L^2$.
The longest overall time is the one needed to flip global magnetisation,
which requires jumping the highest energy barrier $t_{pass} \sim e^{cL^{d-1}}$.
For large $L$, there is an ample regime  $t^*\ll t\ll t_{pass}$ where there
are two well-defined states.
 Many examples for which in fact this construction is the most interesting have
no  parameter that controls the separation of $t_{pass}$ and
$t^*$: the time for relaxation within a metastable 
state, and the time for escaping it, are different but not infinitely so.

\section{Transition currents}

\begin{figure}
\begin{center}
\includegraphics[width=12cm]{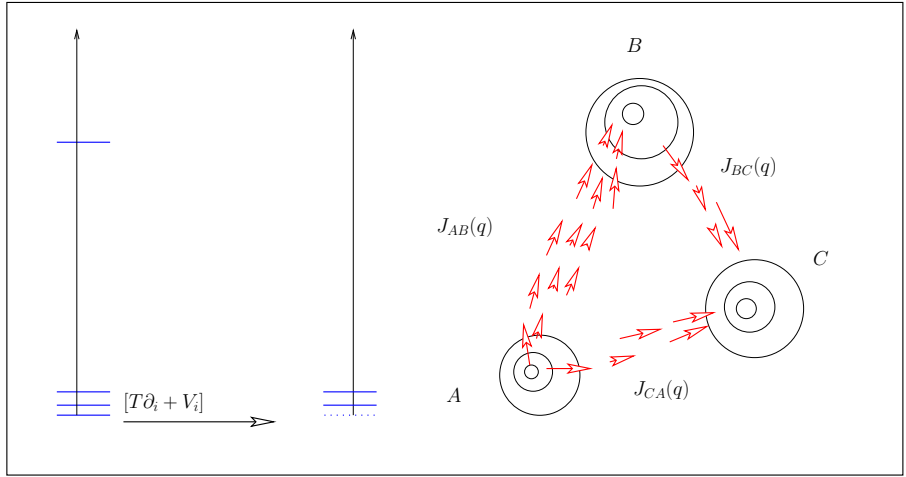}
\end{center}
\caption{Spectrum and escape currents. Acting with 
$\left[T \frac{\partial}{\partial q_i}+\frac{\partial V}{\partial q_i} \right]$ on the eigenvectors 
`below the gap', one obtains a basis for all the interstate currents.} 
\label{current1}
\end{figure}

Given a state metastable state $|P\rangle $ 
constructed as above,
we can find the probability current `leaking' from it directly. For example,
in a Fokker-Planck case   putting  $P({\bf q})=\langle {\bf q }|P\rangle $ :
\begin{equation}
J_i({\bf q}) \propto \left(T\frac{\partial}{\partial
  q_i}+\frac{\partial V}{\partial
  q_i}+f_i\right)P({\bf q})\label{311}
\end{equation}
Because every metastable  state $P$ is a  linear combinations of the
exact eigenvectors
 $\psi^{R}_\alpha({\bf q})=\langle \psi^{R}_\alpha|{\bf q}\rangle$
with eigenvalue below the gap:
\begin{equation}
 P({\bf q}) = \sum_{\alpha=1}^p c_\alpha \psi^{ R}_\alpha({\bf q})
\label{312}
\end{equation}
the associated escape currents are (see Fig \ref{current1})
linear combinations of currents obtained by acting on each of  them 
\begin{equation}
J_i^\alpha \equiv 
\left(T\frac{\partial}{\partial q_i}+\frac{\partial V}{\partial
  q_i}+f_i\right) \psi^{ R}_\alpha({\bf q})
\end{equation}
If the stationary state has no current, then
there are only $p-1$ independent currents, rather than $p$ of them.
  
In many physical situations, we have some idea of the reaction path followed by 
the current, but we do not know the intensity of such a current or the reaction rate.
Let us derive a formula for the rate in terms of an unnormalised current distribution ${\bf J}(q)$.
The only assumption we make is that the reaction time is slower then any other relaxation~\cite{sorin}.
Suppose one has the current ${\bf J}$ 
 escaping a metastable state in an overdamped Langevin problem
$P({\bf q})$ with only conservative forces :
\begin{eqnarray}
J_i({\bf q}) &\propto& \left(T\frac{\partial}{\partial
  q_i}+\frac{\partial V}{\partial q_i} \right)P({\bf q}) \;\;\; with
\;\;\; P({\bf q}) = \sum_{\alpha=1}^p c_\alpha \psi^{ R}_\alpha({\bf q})
 \\ & & \;\; and \;\;\; \int d^N{\bf q} \;
\psi^{L}_\alpha({\bf q}) \psi^{ R}_\delta({\bf q})= \delta_{\alpha
  \delta} = \int d^N{\bf q} \; e^{\beta V} \;
\psi^{R}_\alpha({\bf q}) \psi^{ R}_\delta({\bf q}) 
\label{pipo}
\end{eqnarray}
The last equality in equation (\ref{pipo}) is deduced from the relation between right and left eigenvectors
(\ref{since}). 
We first compute:
\begin{eqnarray}
\int d^N\!\!q \; e^{\beta V} \; {{\bf J}}^2 &=&\int  d^N q 
\left\{\left(T\frac{\partial}{\partial q_i}+\frac{\partial V}{\partial
  q_i} \right)P \right\} \; e^{\beta V} \; \left\{
\left(T\frac{\partial}{\partial q_i}+\frac{\partial V}{\partial q_i}
\right)P \right\} \nonumber \\ &=&T \int d^N\!\!q \; \left\{
\left(T\frac{\partial}{\partial q_i}+\frac{\partial V}{\partial q_i}
\right)P \right\} \; \frac{\partial}{\partial q_i}\left(e^{\beta
  V}P\right) \nonumber \\ &=& T \int d^N\!\!q \; P \; e^{\beta V} \;
(H_{FP}P) = \sum_{\alpha=1} \lambda_\alpha c^2_\alpha,
\end{eqnarray}
and similarly:
\begin{equation}
\int d^N\!\!q\; e^{\beta V}\;(\mbox{div}\; {\bf J})^2 = \int d^N\!\!q
\,(H_{FP}P)\;e^{\beta V}\;(H_{FP}P)
=\sum_{\alpha=1} \lambda_\alpha^2 c_\alpha^2,
\label{summ}
\end{equation}
where we have used the eigenvalue equation and the normalisation (\ref{pipo}).
Let us now assume for simplicity there is only one metastable state, so
that there are $p=2$ eigenvalues `below the gap'.  At large times
$t^*\ll t\ll t_{pass}\sim \lambda_m^{-1}$, only the first non-zero
eigenvalue $\lambda_m$ contributes to the sums, and we get:
\begin{equation}
t_{activ}=\lambda_{m}^{-1} = \frac{\sum \lambda_\alpha c_\alpha^2}{\sum \lambda_\delta^2 c_\delta^2}=
\frac{\int d^N\!\!x \; e^{\beta V} 
\; {{\bf J}}^2}{T\; \int d^N\!\!q \; e^{\beta V}\;(\mbox{div}\; {\bf J})^2}.
\label{timescale}
\end{equation}
Note that the normalisation of the current is irrelevant.
{\em This formula is valid on the assumption of separation of timescales,
irrespective of its cause.}


 For the Kramers equation, a  similar expression can be obtained
in the same way, in terms of  the {\bf reduced current} \ref{reduced}.
A tedious but straightforward calculation yields~\cite{sorin}:
\begin{equation}
t_{activ}={\mbox {Re}} \; \lambda_{max
}^{-1} = \frac{
\gamma \int d^N\!{\bf q}d^N\!{\bf p} \; 
e^{\beta E} \;[\sum_i  J^{red}_{q_i}({\bf q,p}) J^{red}_{q_i}({\bf q,-p})]}
{
 T \int  d^N\!{\bf q}d^N\!{\bf p} \;
e^{\beta E}\;
{\mbox {div}}_{\bf q,p} [J^{red}({\bf q,p})]  
{\mbox{div}}_{\bf q,p} [J^{red}({\bf q,-p})]
} .
\label{timescale1}
\end{equation}
note the sum in the numerator runs on coordinates and not on
momenta.  This formulas are useful because they do not depend on a
global normalisation of the current.

\subsection{Arrhenius formula}

The Arrhenius expression for the activated passage at low temperatures
case can be easily derived from the formula (\ref{timescale}), using
the fact that the numerator is dominated the neighbourhood of the 
barrier top,  and the numerator by the neighbourhood of the 
bottom of the starting well.
 \begin{figure}[htbp]
\begin{center}
\includegraphics[width=8cm]{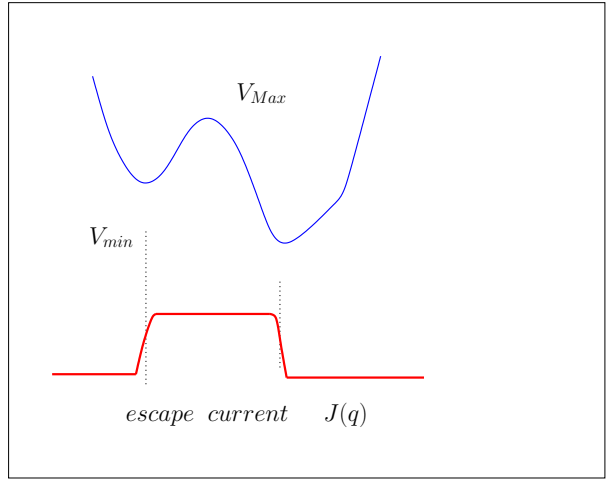} 
\end{center}
\caption{Escape from a metastable state of height $V_{min}$ through a barrier 
of height $V_{Max}$. Below: the current distribution.}
\label{double}
\end{figure}
Let us  show how this works for a one dimensional double well as in Fig. 
\ref{double}. The current starts in the metastable state, is approximately constant,
and falls in the stable state. 

The   divergence of the current $J(q)$ of a state $\psi^R(q)$ satisfies
\begin{equation}
{\mbox{div}} J =H_{FP} \psi(q) =\lambda_m \psi(q)
\end{equation} 
where $\lambda_m$ is the first non-zero eigenvalue.
Now, we have seen that if  $\psi(q)$ is a metastable
equilibrium distribution in the departure state, it is proportional
there to the Gibbs-Boltzmann distribution {\em in that basin}, so that: 
\begin{equation}
{\mbox{div}} J \propto \psi(q) \propto e^{-\beta V} \;\;\;  
\end{equation}
up to a normalisation.  
The current in a point $q$ within the metastable basin is then:
\begin{equation}
J(q) = \int_{-\infty}^q dq' {\mbox{div}} J(q') =   \int_{-\infty}^q dq' \;
e^{-\beta V} \sim e^{-\beta V_{min}}   \int_{-\infty}^q dq' \; e^{-\beta V^{''}_{min} q^{'2}/2}
  \end{equation}
where we have used a Gaussian approximation. Around the barrier, we get: 
\begin{equation}
 J_{barrier} \sim  \left(\frac{2 \pi}{\beta V^{''}_{min}} \right)^{1/2} \; 
e^{-\beta V_{min}}
\end{equation} 
The denominator of (\ref{timescale}) is dominated by the neighbourhood of the minimum:
\begin{equation}
\int_{basin} dq \;
e^{\beta V}\;(\mbox{div}\; { J})^2 \sim  \int dq \; e^{-\beta V} \sim 
e^{-\beta V_{min}}  \left(\frac{2 \pi}{\beta V^{''}_{min}} \right)^{1/2} 
\end{equation}
The numerator in (\ref{timescale}) becomes, using the fact that $J$ 
is essentially constant around the maximum
\begin{equation}
\int_{barrier}  dq \;
e^{\beta V}\; { J}^2 \sim J_{barrier}^2  \int_{barrier}  dq \;
e^{\beta V} \sim e^{\beta V{Max}} 
\left(\frac{2 \pi}{\beta V^{''}_{Max}}\right)^{1/2}   J_{barrier}^2
\end{equation}

Putting numerator and denominator together, we get:
\begin{equation}
t_{activ}=\lambda_{max }^{-1} = \frac{\int dq \; e^{\beta V} \; {{
J}}^2} {T\int dq \; e^{\beta V}\;(\mbox{div}\; { J})^2} \sim 2 \pi  \;
\frac{e^{\beta(V_{Max}-V{min})}} {\sqrt{|V^{''}_{Max}V^{''}_{min}| }}
\label{Arrhenius}
\end{equation}
which is Arrhenius formula~\cite{Hanggi}, with the good prefactor. 
Note how simple the argument is, once we have (\ref{timescale}).

\section{Hydrodynamic Limit and Macroscopic Fluctuations}

\begin{figure}[htbp]
\begin{center}
\includegraphics[width=9cm]{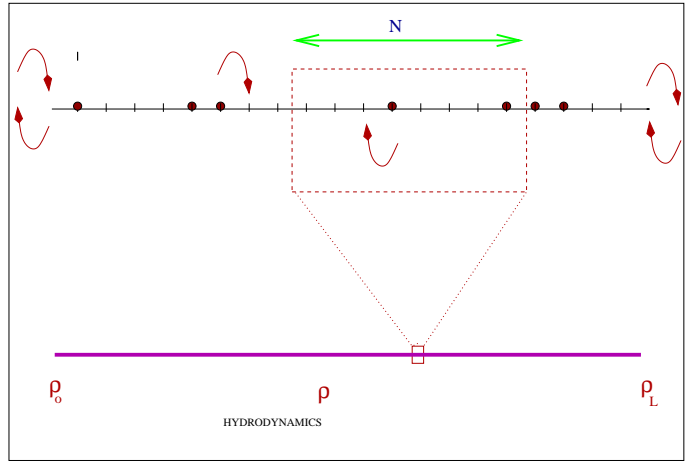}
\end{center}
\caption{ The simple symmetric exclusion process, and its 
coarse-grained hydrodynamic limit.}
\label{sep1}
\end{figure}

The Simple Symmetric Exclusion Process of Equations
(\ref{SSEP}) and Figure \ref{sep1} has a
separation of timescales, this time brought about by a local
conservation law (of particle number) rather than by barriers.
Consider an  $L$-site long, isolated chain, with average occupation one half.
Suppose now that we make a vacancy of twenty contiguous unoccupied sites.
For this to happen spontaneously is an extremely rare event ($~\sim 2^{-20}$),
 and left on its own, the vacancy will be covered rapidly,
in a time of order one. 
On the other extreme, consider a slowly varying average density profile,
say as a sinusoidal oscillation of length $L$. Such a fluctuation will
take a time of order $L^2$ to die out. 

This separation of timescales manifests itself in the stochastic evolution
operator, which has eigenvalues of order $1$  for the most
steep and $L^{-2}$ for the smoothest spatial fluctuations, respectively.
 
We may now choose to study only the smooth fluctuations, corresponding
to the lowest eigenvalues, see~\cite{Spohn1983,Bertini1}. To do this
 we introduce  a parametrisation of space~\footnote{ Note that $x$ here is an index labelling the field
$\rho(x)$. Comparing with the examples in the first sections, we should make
the correspondence $(q_i,i) \rightarrow (\rho(x),x)$}
 $x_k=\frac{k}L$ and rescale the time as $t \to L^2 t$. In the rescaled time,
steep fluctuations disappear immediately. At
the macroscopic level, the density profiles we consider 
are smooth functions and
discrete gradients can be replaced by continuous ones:
\begin{equation}
  \rho_{k+1}-\rho_k \to \frac {\nabla \rho(x_k)}L,\;\;\;
  \hat\rho_{k+1}-\hat\rho_k \to \frac {\nabla \hat\rho(x_k)}L, \;\;\;
  \frac 1 L\sum_{k=1}^{L-1}\to \int_0^1 \; d x 
\end{equation}
It can be shown that in terms of these variables,
the evolution of the smooth density fluctuations  is given by
  the stochastic equation~\cite{Spohn1983}
\begin{equation}
  \label{stochdyn}
  \dot \rho  = \frac{1}{2} \nabla^2
  \rho + \nabla[\sqrt{\rho(1-\rho)} \eta] ;\qquad \rho(0)=\rho_0;\qquad \rho(1)=\rho_1
\end{equation}
where $\eta$ is a white noise of variance $1/(2\,L)$.
 This is the  formula for the
fluctuating hydrodynamics of the exclusion process~\cite{Spohn1983}.

We can now construct the action
with the  Martin-Siggia-Rose~\cite{MSR}
formalism. 
Performing the usual steps as in section 1.3, introducing 
conjugate fields  $\hat\rho(x,t)$, we get:
\begin{equation} 
  P({\bf \rho^f},t_f;{\bf \rho^i},0) =
\int D[\rho] \, D [\hat\rho] \; 
e^{
-2L \{   \int_0^{t_f} \int_0^1 \; d x \; d t \; 
(\hat\rho \dot \rho - \tilde H      ) 
\} 
}
  \label{flucthydro}
 \end{equation}
This time, $L^{-1}$ plays  the role of temperature -- or of $\hbar$,
and we have  the `classical' Hamiltonian  density:
\begin{equation} 
\tilde H  = \frac{1}{2} \int \; dx \; \left[ 
\rho(1-\rho) (\nabla \hat\rho)^2
-  
\nabla \hat\rho \nabla \rho
\right]
 \end{equation}
 The
paths are constrained to be $\rho_i(x)$ and $\rho_f(x)$ at initial and
final times, respectively. The values of $\hat \rho$ are
unconstrained, which is in agreement with the fact that this is a
Hamiltonian problem with two sets of boundary conditions.

The message of this section is that  
{\em we may transfer all the `semiclassical' low-noise (small $T$) techniques
to the coarse-grained limit} to obtain a `Macroscopic Fluctuation Theory' 
\cite{Bertini1,Jordan}.

\chapter{Large Deviations}

In equilibrium statistical mechanics we are given the probability of
 being in any particular configuration. For a dynamical system, 
we may wish to ask similar questions concerning {\em histories}, rather than configurations: what is the probability
that the system  visits a sequence of configurations at  given times, or that during a time-interval
its average enery has a given value, and so on. Often   these events are {\em rare}, their probability
is small. In spite of this they may be  important: for example, what  is interesting in chemistry
are reactions that are slow compared to thermal vibrations.
In this section we shall study two types of large deviations: those that are rare because they are induced by (weak) noise,
and those that are rare because they are sustained for a long time. Technically, this provides
us with two small parameters: noise intensity and inverse time-span, respectively.  

\section{Climbing to unusual heights}

We now study  the probability of finding the
system in unusual configurations. The formalism is essentially the WKB
theory for semiclassical quantum mechanics, mathematicians know it
as the Freidlin-Wentzell~\cite{Freidlin} formalism.
 For simplicity, the discussion in this section will be for the overdamped 
case, but one can do the same for any other stochastic equation.

In the small-noise limit, whichever its origin (low temperature, 
hydrodynamic limit), the equations of motion are essentially deterministic.
If we ask for the probability $P({\bf q_o},t_o \to {\bf q},t)$ 
of the system meeting an appointment at time $t$ 
 in ${\bf q}$, given that it started in
${\bf q_o}$ at time $t_o$ we may get two sorts of answers: 
{\em i)} Essentially 
{\em one} if a  deterministic path 
precisely passes by the given points at the given times.
{\em ii)}  Exponentially small
$\sim e^{-\frac{1}{T} {\mathcal{F}( {\bf q},t ) }  }$ otherwise: 
only thanks to the noise the system can get out of its deterministic
schedule in order  meet the appointment 
(this includes being in the right place at the wrong time).
In order to calculate probabilities, we go back to the path-integral expression
(\ref{eqq})
\begin{equation}
P({\bf q_o},t_o \to {\bf q},t)=
\int D[ {\bf q}, {\bf \hat q}] \; e^{\frac{1}{T} \int dt \; [ \sum_i
\hat q_i \dot q_i - {\tilde H}]}
\end{equation}
As we mentioned in section 1, this is an imaginary-time path-integral
with the `classical' Hamiltonian 
\begin{equation}
\tilde H =  -\sum \hat q_i \left( \hat q_i +\frac{\partial V} {\partial q_i}
+f_i\right)
\end{equation}
The path integral is dominated by the extremal trajectories, which 
satisfy Hamilton's equations:
\begin{equation} \left\{
\begin{array}{ccccc}
\dot q_i &=& \frac {\partial H} {\partial \hat q_i} &=& -\left( \hat q_i
+\frac{\partial V} {\partial q_i} +f_i\right) -\hat q_i \\
 \dot{\hat q_i} &=&-\frac{\partial H} { \partial  q_i} &=& 
\sum_j \left(\frac{\partial^2 V} { \partial q_j \partial q_i} \right)
\hat q_j +
\frac{\partial f_j} {\partial  q_i} \hat q_j
\end{array}
\right.
\label{equatio}
\end{equation}
The probability now takes the large deviation form:
\begin{equation}
\ln \left\{ P({\bf q_o},t_o \to {\bf q},t) \right\}
= -\frac{1}{T} \mathcal{F}_t = -\frac{1}{T} {\bf {\mbox{Action}}}
\label{acc}
\end{equation}
which defines the large deviation  function $\mathcal{F}_t ({\bf q}) $.
The action is the integral 
\begin{equation}
 {\bf {\mbox{Action}}} = -\int 
 dt \; [ \sum_i
\hat q_i \dot q_i - {\tilde H({\bf q},{\bf \hat q})}]
\end{equation}
with $({\bf q},{\bf \hat q})$ solution of (\ref{equatio}).

\vspace{.1cm}

{\bf Noiseless solution.}

\vspace{.1cm}

A family of solutions of (\ref{equatio}) can be easily found:
\begin{equation}
\hat q_i=0 \;\;\;\;\;  ; \;\;\; \;\; 
\dot q_i= 
-\frac{\partial V} {\partial q_i} -f_i \;\;\; 
\end{equation}
The dynamics of the original (hat-less) variables is just the noiseless
 equation of motion (\ref{lange2}).
 The action is zero, as can be easily checked:
this means that the large deviation function will be also zero -- in fact,
its smallest  value.

\vspace{.1cm}

{\bf Other solutions.}

\vspace{.1cm}

Other solutions of (\ref{equatio})
can be found with $\hat q_i \neq 0$. They do not correspond
to motion in the original force field (\ref{lange2}), signalling the fact 
that noise is playing an important role. The action is positive,
a fact that can be best appreciated in the Lagrangian formalism (\ref{lag}).
The large deviation function is now positive, and the probability
is exponentially suppressed in $1/T$, again an indication 
  that noise is playing a role.

\vspace{.1cm}

{\bf Large times}

\vspace{.1cm}

In most cases, we are interested in the long time limit
${\mathcal F}_t$
at large times:
\begin{equation}
e^{-\frac{1}{T} F({\bf q})} = \lim_{t \to \infty} 
P({\bf q_o},t_o \to {\bf q},t)
\end{equation}
or, otherwise stated, in the probability
of finding a stationary  system  in a configuration ${\bf q}$. 
The corresponding large-time 
deviation function  (which we shall denote simply as $\mathcal F$)
becomes independent of the initial condition. How can this be?

\begin{figure}
\begin{center}
\includegraphics[width=12cm]{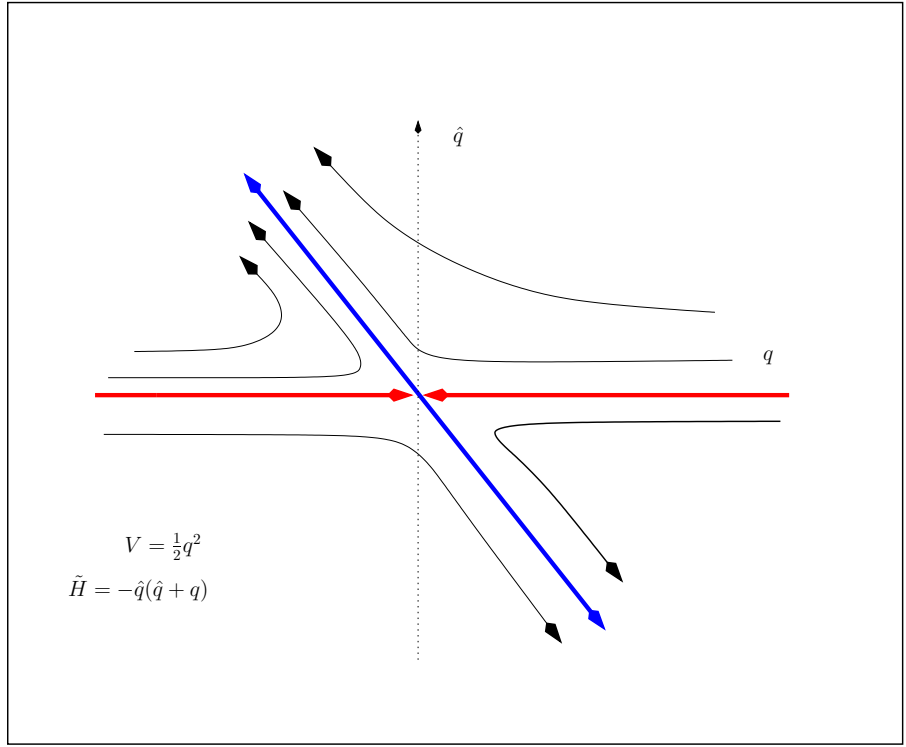}
\caption{The $(q,\hat q)$ space for a Langevin process in a one-dimensional
quadratic potential. The incoming straight  lines indicate the noiseless 
`downhill'
trajectories,
the outgoing straight lines the `uphill' trajectories.
 Curved trajectories missing the origin are  
relevant for finite-time large deviations.}
\end{center}
\label{harmo}
\end{figure}

Consider first a quadratic potential $V=\frac{1}{2} q^2$
in one dimension, as in Fig. \ref{harmo}. The Hamiltonian is 
$\tilde H = -\hat q (\hat q + q)$. The `classical' trajectories 
are given by:
\begin{equation} \left\{
\begin{array}{ccccc}
\dot q &=& -2 \hat q - q \\
 \dot{\hat q} &=& 
\hat q 
\end{array}
\right.
\label{equatio2}
\end{equation}
The solution with $\hat q(t)=0$ corresponds to the relaxation to the minimum.
If we now start at time $t_o$ at, say, $q=1$ and ask  that at time $t>t_o$
we be at $q=-1$,  we have to 
take  one of the  trajectories with $\hat q > 0$ ;
 the one that arrives in $q=-1$ at the right time. 
If we now consider very large times $t$, the solution will be one that
passes close to the  (hyperbolic) point $(q=0 \, , \, \hat q =0)$, in the vicinity of which
it will spend a long time. 
In the limit $t \rightarrow \infty$, the trajectory is the succession of a 
 gradient descent into the origin, and an `uphill'  trajectory emerging
from the origin.
Similarly, the trajectory that starts in $q=1$ and ends at large
times at $q=+2$  is composed of a gradient descent towards
 the {\em left}, followed
by an uphill motion to the {\em right} -- the limit of a $\hat q<0$ `bounce'.

In conclusion, in order  to calculate  the stationary
large-deviation function we have to consider the `downhill' trajectory 
(the  {\em anti-instanton}) with $\hat q=0$ from the initial to
the stationary point, followed by an `uphill' ({\em instanton}) trajectory
from the stationary point to  the final point.
\begin{figure}
\begin{center}
\includegraphics[width=12cm]{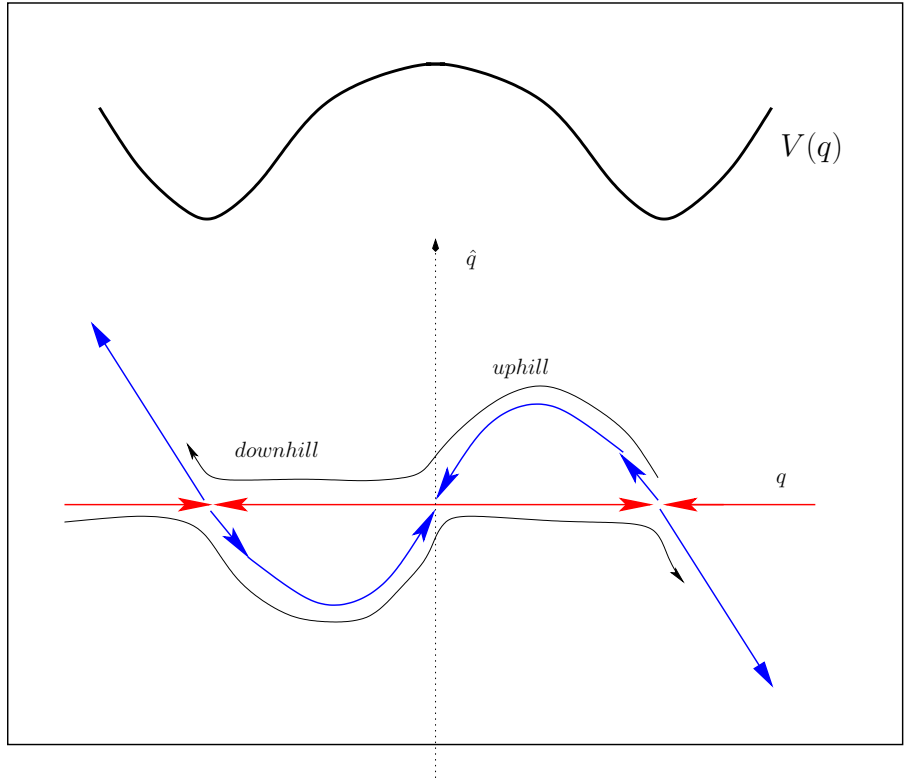}
\caption{Trajectories in $(q,\hat q)$ space for a double-well potential.}
\label{dw}
\end{center}
\end{figure}
The large deviation function is the sum of the downhill  and
the uphill actions. As mentioned above, the former is zero: When calculating 
the probability of reaching a
 configuration
at large times, we may consider that we started from a stationary point -
a physically intuitive result since in a short time at the beginning
the system goes from initial to stationary configuration.

We need the `uphill', instanton trajectories
to calculate, via their action, the probability of a rare configuration.
 {\em For  generic dynamics this is hard problem to solve
analytically, and even numerically}.

\subsection{Detailed balance and Onsager-Machlup symmetry}

As we saw in section 2, if there is detailed balance, and
given that one trajectory dominates for the downhill process,
  we should expect the time-reversed 
trajectory to dominate for the uphill process.
Let us check this  explicitly for a multidimensional
system with a potential $V$  and no forcing $f_i=0$. We propose,
as partial  solution of (\ref{equatio})
\begin{equation}
\dot q_i = {\bf +} \frac{\partial V}{\partial q_i}
\label{up}
\end{equation}
Inserting in the first of (\ref{equatio}), this implies $\hat q_i=
-\frac{\partial V}{\partial q_i}$
which, when replaced in the second of (\ref{equatio}) gives:
\begin{equation}
\dot {\hat q_i} = 
\sum_j \frac{\partial^2 V} { \partial q_j \partial q_i} \hat q_j
\end{equation}
Replacing $\hat q_i=
-\frac{\partial V}{\partial q_i}$ in this equation, and using 
\begin{equation}
 \frac{d}{dt} \left(-\frac{\partial V}{\partial q_i}\right)
 =-\sum_j \frac{\partial^2 V} { \partial q_j \partial q_i} \dot q_j
\end{equation}
we obtain an identity. We conclude that the time-reversed dependence~\footnote{
Note that time reversal applies to the $q$, and not the $\hat q$
variables}
(\ref{up}) is indeed a solution. 
We may compute the action of
the uphill trajectory, which turns out to
depend exclusively on the initial and final potentials:
\begin{eqnarray}
{\mbox{action}} &=&  \int dt \;  \sum_i
\underbrace{ {\hat q_i} }_{-\frac{\partial V} {\partial q_i} } \dot q_i+
\int dt \;
\sum_i \hat q_i \underbrace{
\left( \hat q_i +\frac{\partial V} {\partial q_i} \right)}_{=0}
\nonumber \\
&=& \int dt \; \frac{dV} {dt} = V({\bf  q}) - V({\bf q_{min}})
\end{eqnarray}
in accordance with the general situation with detailed balance 
discussed in Section 2.1
Up to a multiplicative constant, this implies via (\ref{acc}) 
that the stationary
probability is the Gibbs-Boltzmann weight, as expected.

On the other hand, repeating the calculation in the case
in which there is 
a generic force term $f_i$, we get that the second
of (\ref{equatio}) is satisfied by the time-reversed trajectory if:
$\sum_j \left( \frac{\partial f_i}{\partial q_j}- 
\frac{\partial f_j}{\partial q_i}
\right) f_j =0$, which is in general not true. Hence,
we  reach the important conclusion that {\em in the presence of 
forces that do not derive from a potential, relaxations into and
excursions out of the stationary state are not the time-reversed of
one another}.
Furthermore, we have to calculate the action
 on the basis of the explicit solution,
and this it is no longer miraculously given exclusively in terms of the
initial and final configurations.

\subsection{The Arrhenius Law again}

In the preceding paragraphs we have assumed that in order to reach
the final configuration, only one downhill and one uphill trajectory
suffice. This is clearly the case when there is only one stable state.
In the presence of metastability, the trajectory may be a sequence
of downhill and uphill segments. 

For example, in the case of a double-well potential, the passage probability 
(the Arrhenius law) can be obtained  as the probability
of falling onto a stable point, then climbing up to an unstable point,
and then  
descending to the next saddle, as in figure \ref{dw}. The probability of descents 
 being of order unity,
we are only left with the uphill path.
Again, if the system derives from a potential,
the probability of climbing depends only on
the difference in height between valley and saddle, and we
recover the Arrhenius law calculated in previous sections
 (\ref{Arrhenius}). In the
presence of non-conservative forces, then
the path joining stable and unstable saddles has to be computed, and
from its action
we get the probability.

\vspace{.2cm}

\fbox{\parbox{12cm} {%
In many dimensional energy landscapes, one may wonder
if the most probable path still goes through
saddles with only one unstable direction.
 A moment's thought shows this to be the case, as discussed by Murrel and 
Laidler \cite{Murrel} 
}}

\vspace{.2cm}

\subsection{Low noise in phase-space: Kramers and thermostatted.}

Let us mention briefly how one proceeds in the case of phase-space
dynamics with noise.
One obtains by following the same steps as in the previous paragraph,
 `classical' equations in an extended space
$(q_i,p_i,\hat q_i, \hat p_i)$. There are solutions of these 
equations that have $\hat q_i=\hat p_i=0$: they correspond
to the original noiseless equations in the original space and have zero action.
The rest of the solutions 
 have non-zero $(\hat q_i,\hat p_i)$
and  positive action.
 
In some cases, one may 
go to a `Lagrangian'~\footnote{Note that $({\bf q},{\bf p})$ are the 
`coordinates', and $({\bf \hat q},{\bf \hat p})$ 
the `momenta' in the $4N$-dimensional phase-space}
 description involving only $(q_i,p_i)$,
but then the equations obtained contain second time-derivatives, a relic of
the noise in the original phase-space.

\vspace{.2cm}

\fbox{\parbox{12cm} {%

{\bf Periodic orbits, complexities and traces}

\vspace{.2cm}

{\bf Classical Mechanics}

\vspace{.2cm}

What follows is very sketchy, its only purpose is to stimulate the curiosity
of the reader, who will find an excellent reference on the subject 
\cite{predrag}.
As we have seen in the previous sections, the spectrum of an
evolution operator contains all the  information on the ergodic properties
of the system. One way to study the spectrum of any operator $H$ is
to compute the trace ${\mbox{ Tr} } e^{-tH}$, and then obtain
the resolvent, which has poles in the eigenvalues $\lambda$ of $H$, through
\begin{equation}
\sum_i \frac{1}{\lambda - \lambda_a} =  {\mbox{ Tr} } [\lambda -H]^{-1}=
\int_0^\infty dt\; {\mbox{ Tr } } e^{-t(\lambda-H)}
\label{period}
\end{equation} 
The trace on the right hand side is a sum over paths just as seen in the
previous sections, the only difference is that we are to consider
{\em closed} trajectories in which initial and final configurations
coincide.  For an evolution taking place in phase-space
$({\bf q},{\bf p})$, 
it is an integral of probabilities of return after time $t$ 
$ {\mbox{ Tr} } e^{-tH} \sim \int d{\bf q}\; d{\bf p} \;
 P({\bf q},{\bf p},t \, ; \, {\bf q},{\bf p},t=0)$.

As mentioned in section 2, one can study the chaoticity 
properties of Hamiltonian dynamics by studying the spectrum of its evolution
operator in the presence of noise, and then letting the 
noise level go to zero: one thus   uncovers  Ruelle-Pollicott resonances.
This poses the problem
that noise will make energy nonconserved, and  generate a slow
diffusion in energy.
To avoid this, we may use the energy-conserving noise of the Gaussian thermostat
in section 1, leading to $H_G$ : 
\begin{equation}
 H_G= \sum_i \left[
\frac{p_i}{m} \frac{\partial}{\partial q_i}-\frac{\partial V({\bf q})}{\partial q_i}
\frac{\partial}{\partial p_i} \right]- \epsilon \sum_{jl} \left[
\frac{\partial}{\partial p_j}
g_{jl} \frac{\partial}{\partial p_l} \right]
\label{eq33}
\end{equation}
cfr. Eq. (\ref{eq3}) with ${\bf f}=0$.

A trace in the path integral becomes then a sum of periodic orbits on
the energy shell. In the small-noise limit those that  dominate are 
the ones  having zero action, and these are just the periodic
orbits of the original Hamilton's equations.  One thus 
expresses  ${\mbox {Tr}} e^{-tH_G}$ as a sum over orbits of period $t$.  

The final product is that Equation (\ref{period}) becomes a relation
between the resolvent, containing the information on eigenvalues,
and a sum of periodic orbits of all periods.
The interested reader will find this properly done in \cite{predrag}.

\vspace{.2cm}

{\bf Complexity in Statistical Mechanics}

\vspace{.2cm}

A different application of the same idea~\cite{biku} is to count the number of metastable states.
With the assumptions of section 3.1, picking a time $t$ intermdiate between the time needed to 
equilibrate within a state $t^*$ , and the time needed to escape it $t_{pass}$,
the number of states is given by ${\mathcal N}_{states} =  {\mbox{ Tr } } e^{-tH}$ with $t^*\ll t \ll t_{pass}$.
 This becomes a sum over all periodic trajectories of period $t$.

}}

\vspace{.2cm}

\fbox{\parbox{12cm} {%

{\bf Sampling transition paths  in practice}

\vspace{.2cm}

A problem of great interest in physics and chemistry is 
the one of computing in practice the escape route (and rate) from a
 metastable state, for example the decay rate of a metastable molecule.
In many cases we have a separation
of timescales between molecular vibration and the time needed for 
the actual decay to occur.
In order to compute the probability $P(a,t;b,t_o)$ 
of starting at $t_o$ in a configuration
$a$ belonging to the metastable state and reaching  a configuration $b$
belonging to the stable state at time $t$, we may  sum
over paths going from $a$ to $b$, with their appropriate weight given by 
the action.

Depending on the dynamics, different approaches are more practical.
In the case of a system that is strongly coupled to
a thermal bath, and follows a Langevin equation, one can follow the Lagrangian 
`polymer' analogy (\ref{lag}), and use any Monte Carlo simulation method
for a system in equilibrium at temperature $T$ to sample the paths.

For a system that is closer to being deterministic, a well-developed technique
(TPS: `Transition Path Sampling')~\cite{Dellago} uses 
an algorithm that samples  trajectories by modifying them slightly  
at the barrier. This 
kind of change allows to obtain  a new path that still has good chances
of being a transition (going from state $a$ to state $b$), at least
if the system is not too chaotic.

There is a large body of literature on the subject (see e.g. 
\cite{Dellago,Eijnden,parrinello})  since the potential
applications in chemistry, biochemistry and physics are huge.
}}
\vspace{.2cm}

\section{Unusual time averages}

In this section we study a different type of large deviation.
Instead of asking for the probability of the system reaching an unusual place,
we ask for the probability that it sustains an unusual time-average
 for an observable during a long interval:
\begin{equation}
\frac{1}{t} \int dt' \; A(t') = \overline A
\end{equation}
The most celebrated example is the average power, which can be interpreted
in some cases as entropy production:
\begin{equation}
\sigma_t= - \frac{1}{t} \int_0^t dt' \; \frac{\bf f \dot v}{T}
\label{sigm} 
\end{equation}
where ${\bf v}={\bf \dot q}$
Another example is the average potential
energy 
\begin{equation}
\overline V = \frac{1}{t} \int_0^t dt' \; V(t')
\label{pote} 
\end{equation}
Expressing the probability in terms of trajectories:
\begin{equation}
P (A)  = 
\sum_{\mbox{Trajec.}} \;\left( {\bf \mbox{Prob. Trajectory}}\right)
\delta\left(t\overline{A}-\int_o^t \; dt' \; A(t')\right)
\label{dela} 
\end{equation}
and writing  the delta function as an exponential
\begin{equation}
 \delta\left(t\overline{A}-\int_o^t \; dt' \; A(t')\right) = \int_{-i
 \infty}^{+i\infty} d\mu \;e^{\mu \left(t\overline{A}-\int_o^t \; dt'
 \; A(t')\right)}
 \end{equation}
we get:
\begin{equation}
\begin{array}{ccccc}
P(A)  &=& \int_{-i \infty}^{+i\infty} d\mu \; 
e^{\mu t\overline{A}} & &\underbrace{
\int D[{\bf q}] \;\left( {\mbox{Prob. Trajectory}}\right) 
e^{-\mu \int_o^t \; dt' \; A(t')}  } \nonumber \\
& & \;\;   & &  \;\;\;\;\; \downarrow \nonumber\\
&=&  \int_{-i \infty}^{+i\infty} d\mu \; e^{\mu t\overline{A}} & & \;\; \times \;\; 
e^{-tG(\mu)}
\end{array}
\label{cdb}
\end{equation}
which defines the large-deviation function  $G(\mu)$.
For
example, in the Fokker-Planck case, it reads:
\begin{equation}
e^{-tG(\mu)}= \int D[{\bf q}, {\bf \hat q}] \; \; 
e^{\frac{1}{T} \int dt' \; 
\left[\sum_i
\hat q_i \dot q_i - { \tilde H_{FP}} -\mu \int_o^t \; dt' \; A(t') 
\right] }
\label{cda}  
\end{equation}

What we have done is nothing but the analogue of a 
passage from a microcanonical
calculation of `entropy' $= \ln \bar A$, to a canonical calculation
of `free energy' $G(\mu)/\mu$ at `inverse temperature' $\mu$. The `space' in
our problem is in fact the time, and `extensive quantities' are those
that are proportional to time:  we extracted a time in the definition
of $G(\mu)$ in order to make it `intensive'.
For large $t$, in analogy with the thermodynamic limit, assuming that
$G(\mu)$ has a good limit, we may evaluate the integral over $\mu$ by saddle
point, to obtain:
\begin{equation}
\ln P(A)  \sim t[\mu^* A -G(\mu^*)]
\end{equation}
 with
\begin{equation}
A(\mu^*)=\left. \frac{dG}{d\mu}\right|_{\mu^*}
\label{energylike}
\end{equation}
This is the Legendre transform taking from canonical to microcanonical.
Note that $A(\mu^*)$ plays the role of $E(\beta)$ in ordinary 
thermodynamic systems.
Now, by simple comparison, equation (\ref{cda}) can be brought back to
operator language:
\begin{equation}
e^{-tG(\mu)} = \langle {{final}}|e^{-t[H_{FP}+\mu A({\bf
q})]}|{ init }\rangle
\label{cee}
\end{equation}
A similar expression holds for Kramers equation $H_K$.
In the language of the analogy with one-dimensional models, this is just
the transfer matrix formalism~\cite{Huang}, along the time-dimension.
If we now introduce the right and left eigenvectors of   $H+\mu A$:
\begin{equation}
\left[H+\mu A({\bf q})\right]|\psi^R_i(\mu)\rangle = \lambda_i(\mu)
|\psi^R_i(\mu)\rangle \;\;\;\; ; \;\;\;\; \langle |\psi^L_i(\mu)
|\left[H+\mu A({\bf q})\right] = \lambda_i(\mu) \langle |\psi^L_i(\mu)
| \nonumber
\label{express}
\end{equation}
we obtain:
\begin{equation}
e^{-tG(\mu)} =\sum_a \;  \langle {\mbox{final}}|\psi^R_a(\mu)\rangle 
\langle \psi^L_a(\mu) 
|{ init }\rangle \; e^{-t\lambda_a(\mu)}
\end{equation}
and, as $t \rightarrow \infty$:
\begin{equation}
G(\mu)= \lambda_{min}(\mu)
\end{equation}
where $\lambda_{min}(\mu)$ is the eigenvalue with the smallest real part.
Again, this is in complete analogy with the expression of free-energy
densities in terms of the transfer-matrix lowest eigenvalue.

{\em Note the fundamental difference between this large-deviation functions and
those of section 4.1}: in that case by `large' deviations we meant that
they are exponentially small in the temperature (or the coarse-graining size) 
  while here we mean that they are sustained for long times,
and the only large parameter  is precisely the time $t$.

\section{Simulating large deviations (and Quantum Mechanics)}

Large deviations can be measured by evolving the system with its real
dynamics, and then making a histogram of the deviations obtained.
However, equation (\ref{cee}) suggests that we try to simulate directly a
system that evolves through $H_{FP}+\mu A$ ~\cite{GKP,Grassberger}:
\begin{equation}
\dot P = -[H_{FP} + \mu A] P \;\;\; \rightarrow P({\bf q},t)=
e^{-t[H_{FP} + \mu A]}
\label{clo}
\end{equation}
Clearly, as $t \rightarrow \infty$ the distribution $P$ tends to the 
eigenvalue with the lowest real part.

 We are
dealing with a dynamics without probability 
conservation. In fact, we can reproduce it by using a large number of 
non-interacting walkers,
 each performing   the original (Langevin) dynamics with
independent noises, occasionally  giving birth to another
walker starting in the same place, or  dying. A negative (positive) 
value of $\mu A({\bf q})$ gives a probability   $|\mu A({\bf q})| dt$ 
 of making a clone or of   dying, respectively, in a time-interval $dt$.
At each time, the global number of clones $M(t)$ changes, in such a way that
for long times $M(t)/M(t=0)\sim e^{-\lambda_{min}(\mu)t}$. In practice,
one can normalise the total number periodically by
cloning or decimating all walkers with a random factor.
 The factor needed to keep the population constant is, again, the exponential
of the lowest eigenvalue.

Notice that imaginary-time Shr\"odinger equation is precisely of
the form (\ref{clo}), with no drift in the Langevin
process and $\mu A$ the quantum potential. 
Indeed, the method described above was developed for precisely
this case, and is called `Diffusion Monte Carlo' (DMC)~\cite{DMC}.

\chapter{Metastability and dynamical phase transitions}

In several places above we have pointed out that the stochastic dynamics can be
seen as a kind of `thermodynamics in space-time'. Trajectories contribute
with a weight  given by the  (Onsager-Machlup) action, much as energy
determines  the Gibbs-Boltzmann weight in thermodynamics.
Large deviation theory 
just  consists  of biasing the measure with
 an extra weight added to the dynamic action and computing the new sum,
which then looses the meaning of a transition probability and becomes a 
large-deviation function. 

Systems with non-trivial dynamical properties sometimes show very little
in their static (time-independent) structure. The typical example is that of glasses,
which are virtually indistinguishable from liquids
from the point of view of organisation of the molecules,
 until one looks at their dynamics, which is dramatically slower.
This situation has motivated some researchers~\cite{ChGa1}
 to look into space-time
thermodynamics - the statistical properties of trajectories - for the missing
structure.
One considers the large deviation theory of 
   systems that are dynamically non-trivial, and indeed  it turns 
out that one often finds ~\cite{Fred} that there is a rich
structure of `dynamic' phase 
transitions  in the large-deviation functions. 

In this section we shall see that space-time transitions are closely related to
the approach to metastability of section 3, to which
 they provide  useful insights.

\section{A simple example}

\begin{figure}
\begin{center}
\includegraphics[width=14cm,height=8cm]{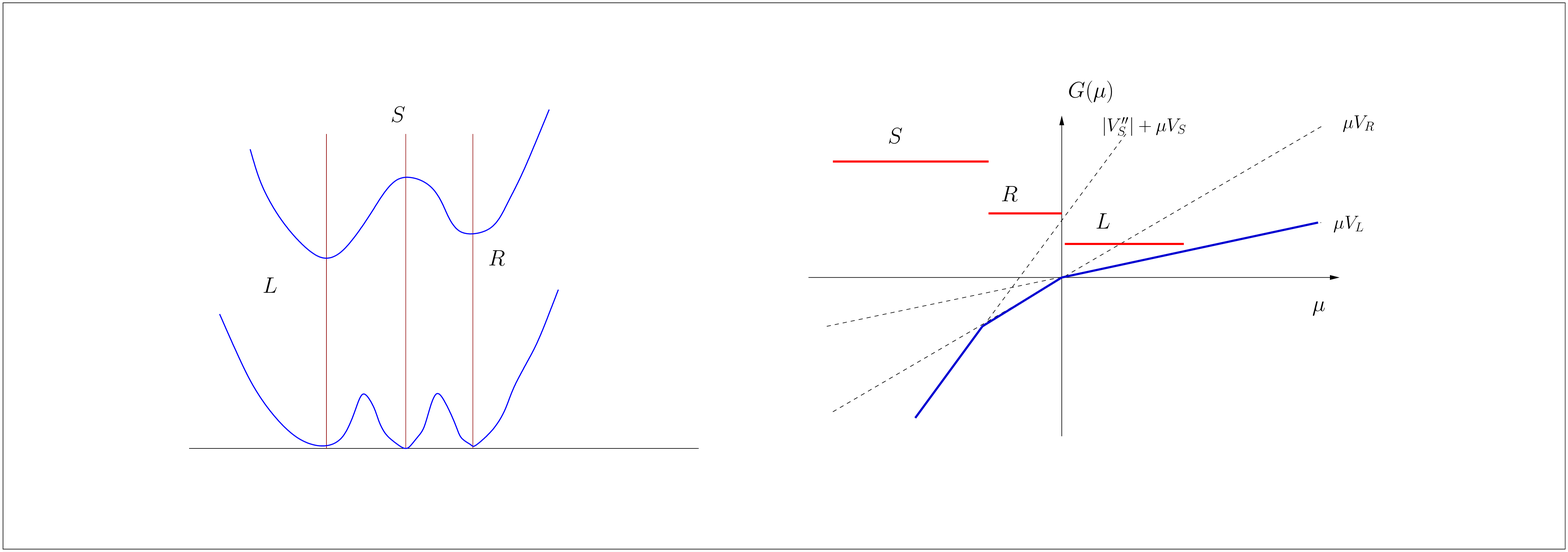}
\caption{Left: the potential $V$ and
 the related effective potential $V_{eff}$. Right: energy (horizontal lines)
and the associated large-deviation function $G(\mu)$. The system has two
first order transitions at $\mu=0$ and at $\mu=-(V_s-V_r)/|V_s^{''}|$ }
\end{center}
\label{largede}
\end{figure}

Let us start by a simple example of a particle in a potential $V$ performing
overdamped Langevin motion at low temperature $T$ (see Fig. \ref{largede}).  
We shall consider the large deviation function of  energy $G(\mu)$ ,
associated with the probability of observing an average energy $\bar V$
(\ref{pote}) 
over long time-intervals. As we saw in the previous section, 
$G(\mu)$ is obtained from the lowest eigenvalue of $H_{FP} - \mu V$.
\begin{equation}
H_{FP} + \mu V= \frac{2}{T} \left[ -T \frac{d^2}{dx^2} +
  \frac{{V'}^2}{8} -\frac{T}{4} V'' + \frac{\mu}{2} TV \right]
\label{mnmn}
\end{equation}
(see \cite{BenziPaladin} for a similar application).
The lowest eigenvalue at low temperatures is given by the same
developments as in section 3.1.1, only that now we have to add the extra
term proportional to $\mu$ in (\ref{mnmn}). At small $T$ each  minimum in the effective
potential 
\begin{equation}
V_{eff}(\mu)=  \frac{{V'}^2}{8} -\frac{T}{4} V'' + \frac{\mu}{2} TV 
\end{equation}
contributes separately, just as in section 3.1.1. 
To leading order in $T$ the contribution of saddles point of $V(q^s)$ at $q^s$ 
is $\lambda \sim \mu V(q^s)$ if it is a minimum,\\ and  
 $\lambda \sim [|V''(q^s)|+ \mu V(q^s)]$ if it is a maximum.
 The lowest amongst
all eigenvalues  dominates:
\begin{equation}
G(\mu)=\lambda_{min}= {\mbox{min}} \left\{
\begin{array}{ccc}
\lambda_L&=&\mu V_L \\
\lambda_R&=& \mu V_R \\
\lambda_S &=& |V^{''}_S| + \mu V_S \\
\end{array}
\right.
\end{equation}
The values of $\bar V$   are given by the Legendre transform
 $V(\mu^*)=\frac{dG}{d\mu}$, and read:
\begin{equation}
V(\mu^*)= \left\{
\begin{array}{cccc}
 V_L &\;\;\; &\mu^*>&0\\
 V_R &;\;\; &\mu^*<&0 \;\; and \;\;  \mu^*>-|V_s^{''}|/(V_s-V_r)\\
 V_S &\;\;\; &\mu^*<&-|V_s^{''}|/(V_s-V_r)
\end{array}
\right.
\label{coex}
\end{equation}
there are two first order phase-transitions, see Fig. \ref{largede}.

Let us pause and analyse this physically. The scenario is typical first order,
with three homogeneous `phases' in time, corresponding to the three
values of Equation (\ref{coex}). When we condition a long trajectory to having
a time-averaged energy $V_o$, this is realised by the system by making
a `phase coexistence' of periods $t_L,t_S,t_R$ spent in each of the three 
stationary points, such that $t=t_L+t_S+t_R$ and $V_ot=V_Lt_L+V_St_S+V_Rt_R$.
This is strictly analogous to the ice/water coexistence when the total
energy is fixed.
{\em An important lesson is that if the system is conditioned to having
a value of an observable  intermediate between that which it takes
in  two metastable phases, it prefers to achieve this
by spending 
 some time in each state, rather than all the time in an intermediate
situation}.

\section{Spectral properties and phase transitions}

One can in fact show that the situation we have seen above is very general.
In particular, it is quite normal that we should have a first order
transition at $\mu=0$.
To see this, let us use the formalism of section 3.
Consider the operator $H$. If there are $p$ independent
long-lived states, with eigenvalues $\lambda_a<t^{-1}_{pass}\sim 0$, and a time
$t^*$ to thermalise inside a state, 
we  can construct a basis of $p$ right and left eigenvectors $P_a$, $Q_a$,
with $  
 \langle  Q_a|P_b\rangle \sim \delta_{ab} $
having essentially zero eigenvalue.

Let us now calculate the eigenvalues of $H+\mu A$, 
for small $\mu$, but still $\mu \gg t_{pass}^{-1}$. We may use  
(non-Hermitian) first order perturbation theory, to get:
\begin{equation}
\lambda_a \sim \lambda_a(\mu=0) + 
\mu \langle Q_a| A| P_a\rangle \sim  \mu \langle A \rangle_{in \; state \; a}
\end{equation}
In other words, the quasidegenerate eigenvalues split 
proportionally to the expectation value of the observable  $A$
in each state, and  the phase that dominates is
\begin{equation} 
G_\mu= \min_{\mbox{states } a } \{ \langle A \rangle_a \}
\label{sele}
\end{equation}
Remarkably, the distribution function has in fact pinned down a `pure state'
$P_a$, $Q_a$, using the observable $A$.
When the sign of $\mu$ is reversed, 
the `minimum' in equation (\ref{sele}) is transformed into a `maximum'
and the selected state is changes. There is hence always a first order
phase transition.
Playing with a different observable, we may make a different 
transition that selects any state.
Hence, we conclude that the dynamic phase transition 
approach is in fact equivalent to the metastability one of 
that of section 3, but it gives us new tools and a practical perspectives.

\chapter{Fluctuation Theorems and Jarzynski equality}

Nonequilibrium work relations, the Fluctuation 
Theorem~\cite{ecm,ft_stationary,GC}, 
and Jarzynski's~\cite{Jarzynski}
equality, are very general results  valid for strongly out of equilibrium systems.
They concern the large deviations of work. As such they  are closely related to -- and enrich our perspective of -- 
the second Law of thermodynamics.
The two 
subjects are quite similar, and in fact may in some cases be encompassed
into a single, more general result: Crooks's equality ~\cite{Crooks}, which
we shall not review here.
 These results are very recent -- surprisingly so, given their technical
simplicity, and have received in the last fifteen years enormous attention.
They are both based on the relation (section 2.3) 
between time-reversal symmetry breaking
on one side, and  work and entropy on the other.

\section{The fluctuation theorem(s)}

 We shall consider here only Langevin processes with inertia, and the Kramers
equation. This makes the discussion simpler, because of the fact mentioned
in previous sections that  power, being a product of force times a velocity,
is only a continuous function of time when there is inertia.

We have seen in section 2.3 that the `time-reversal' symmetry
becomes, in the presence of forcing:
\begin{equation}
\left[ \Pi e^{\beta { H}} H_{K} e^{-\beta { H}} \Pi^{-1}
  \right]^\dag =\; H^\dag_{K} + \frac{d(t\sigma_t)}{dt}
\label{oo}
 \end{equation}
The violation of the symmetry is proportional to a quantity
 that  has the form of an entropy 
 production  (see  (\ref{sigm}) and (\ref{wok})):
\begin{equation}
\sigma_t= \frac{power}{T} = - \frac{1}{tT} \int_0^t dt' \; {\bf f. v}
\label{avg}
\end{equation}
Equation (\ref{oo}) is the basis for the results we shall discuss.
In fact, there are several variants of time reversal, and each gives
different identities~\cite{Chetrite}.

\vspace{.2cm}

\fbox{\parbox{12cm} {%
{\bf General: the implications of an explicitly broken symmetry in statistical mechanics.}

\vspace{.2cm}

The fluctuation theorem makes use of the explicit breaking of a
discrete symmetry, the detailed balance relation. In fact, whenever we
have a system composed of a part that is symmetric under a
transformation, plus an {\em anti}-symmetric perturbation, we can
derive a relation for the large deviation of the perturbations.
Consider for example the statistical mechanics of 
a  system with variables $s_1,...,s_N$
and energy $E_o$,  having
the discrete symmetry $ E_o({\bf {s}})=E_o({\bf {-s}}) $. 
This symmetry 
implies the vanishing of all odd correlation spin functions.  Now, let
us perturb the energy with a field 
 $E({\bf {s}})=E_o({\bf {s}})-\frac{h}{2}M({\bf {s}})$,
conjugate to  a term 
$ M({\bf {s}})=\sum_i s_i $ with
$M(-{\bf {s}})= -M({\bf {s}}) $, so that now
\begin{equation}
E(-{\bf {s}})= E({\bf {s}})+hM({\bf {s}})
\label{vi}
\end{equation}
Can we conclude something in the presence of $h \neq 0$, when the
symmetry is explicitly broken? Indeed, we can: consider the
distribution of the symmetry-breaking term
\begin{equation}
P\left[ M({\bf {s}})=-M \right]= \int \; d{\bf{s}}\; 
\delta\left[ M({\bf {s}})+M \right]\;e^{-\beta (E_o-\frac{h}{2}M)}
\end{equation}
Changing variables ${\bf {s}} \rightarrow -{\bf{s}}$, and using the 
symmetry of $E_o$:
\begin{equation}
P\left[ M({\bf{s}})=-M \right]= \int\; d{\bf{s}}\; 
\delta \left[- M({\bf{s}})+M \right]\;e^{-\beta (E_o+\frac{h}M)}=e^{-\beta hM}
P\left [M({\bf{s}})=M \right]
\end{equation}
or
\begin{equation}
\frac{ P\left[ M({\bf{s}})=M \right]}{ P\left [M({\bf{s}})=-M
    \right]}=e^{\beta h M}
\label{MM}
\end{equation}
One wonders if this elementary property has 
 ever been used in other fields of physics before
the fluctuation theorem.
On the other hand, a derivation of the Fluctuation theorem that makes
close contact with this thermodynamic property has been given by Narayan and Dhar~\cite{Narayan}.

}
}

The fluctuation theorem is a statement about the distribution $P(\sigma_t)$
of the average quantity (\ref{avg}), when the experimental 
protocol is repeated many times. It reads:
\begin{equation}
\frac{P(\sigma_t)}{P(-\sigma_t)} \sim e^{t \sigma_t}
\label{FT}
\end{equation}
A relation like this was first proposed by Evans, 
Cohen and Morriss~\cite{ecm}. 

The second Law of Thermodynamics states that the work done on a system {\em over long times}
must be positive. Equation (\ref{FT}) is then a statement about the `violations'  of the Second Law,
when the average work has the opposite sign~\footnote{The quotation marks 
are just to remember that since the Second Law applies to the limit of long times $tN \to \infty$ for a single-instance experiment, 
these are no true violations.}. The factor $t$ in the exponent to a certain extent quantifies
the supression of the probability of such processes when the time is large, thus giving a better perspective of
the Second Law.

Equation (\ref{FT}) can be reexpressed multiplying by $e^{-t\mu \sigma_t}$ and integrating
over $\sigma_t$ as a property of the large-deviation function
$ G(\mu) = \frac{1}{t} \int d\sigma_t \; P(\sigma_t) e^{-\mu \sigma_t t}$
(\ref{cdb},\ref{cda}):
\begin{equation}
G(\mu) = G(1-\mu)
\label{FT1}
\end{equation}

The result  (\ref{FT})-(\ref{FT1}) is 
extremely general~\cite{ecm,GC,ft_stochastic1,ft_stochastic2}, 
independent on the model's parameters,
and valid for several types of dynamics. Two different settings have to be
distinguished:
\begin{itemize}
\item {\bf Transient:} Each measurement of $\sigma_t$ is made starting
 from a thermalised system at temperature $T$, a configuration chosen with
the Gibbs-Boltzmann distribution.
  At time $t=0$,
non-conservative forces are switched on, $\sigma_t$ is proportional to the
 work
 they make  during a time $t$, which need not be long.
 The system may
be isolated or connected to a thermostat, which then has to be
 at the same temperature $T$.
\item {\bf Stationary:} Here, the sampling is of a 
 system that by assumption has achieved stationarity in the
presence of forcing. For this to be possible 
 it needs a thermostat to absorb heat.
The system is  not  in equilibrium, 
 and the fluctuation relation is valid only in the limit of long
  sampling periods $t \rightarrow \infty$.
\end{itemize}
Another distinction we can make is whether the dynamics is 
{\bf stochastic} (e.g. Langevin) or  {\bf deterministic} (the Gaussian
thermostat (\ref{Gau}) without energy-conserving noise).
The only hard case is the {\bf stationary and deterministic} one,
 the Gallavotti-Cohen~\cite{GC} theorem. It is not only technically
more subtle, but it relies on  a real physical condition that the system 
has to meet, as we shall see.

\subsection{Transient, with or without bath}

Equation (\ref{wok}) can be rewritten, for all $\mu$ (for the moment an 
arbitrary number):
\begin{equation}
\left[H_K - \mu \frac{{\bf f.v}}{T} \right]^\dag =
\left( \Pi e^{\beta {\mathcal H} } \right) \left[ H_K - (1-\mu) \frac{ {\bf
    f.v}}{T}\right] \left( e^{-\beta {\mathcal H}} \Pi^{-1} \right) 
\label{che}
\end{equation}
Using the expression (\ref{cee}) for the large deviation function, starting
from the Gibbs-Boltzmann distribution $|GB\rangle$, we compute:
\begin{eqnarray}
e^{-t G(\mu)} &=& \langle - | e^{-t \left[ H_K - \frac{\mu}{T} {\bf
      f.v}\right] } | GB \rangle
= \langle GB | e^{-t \left[ H_K - \frac{\mu}{T} {\bf f.v}\right]^\dag
} | - \rangle \nonumber \\ &=& \langle - | e^{-\beta {\mathcal H}} e^{-t \left[
    H_K - \frac{\mu}{T} {\bf f.v}\right]^\dag } e^{\beta {\mathcal H}} | GB
\rangle \nonumber \\ &=& \langle - | \Pi e^{-t \left[ H_K +
    \frac{(1-\mu)}{T} {\bf f.p}\right] } \Pi^{-1} | GB \rangle
\nonumber \\ &=& \langle - | e^{-t \left[ H_K - \frac{(1-\mu)} {T} {\bf
      f.p}\right] }| GB \rangle
=e^{-tG(1-\mu)}
\end{eqnarray}
i.e. (\ref{FT1}). Note that we did not have to use any assumptions
either on times or on the dynamics.

\subsection{Stationary with bath}

To compute the large-deviation function for long times,
we proceed as in the previous section, and introduce
  eigenvectors and eigenvalues as in (\ref{express}):
\begin{equation}
\left[ H_K - \frac{\mu}{T} {\bf f.v}\right] |\psi^R_i \rangle =
\lambda_i |\psi^R_i \rangle \;\;\; ; \;\;\; \langle \psi_i^L| \left[
  H_K - \frac{\mu}{T} {\bf f.v}\right] = \lambda_i \langle \psi_i^L|
\end{equation}
to get:
\begin{equation}
e^{-tG(\mu)}=  \langle - | e^{-t \left[ H_K - \frac{\mu}{T} {\bf
      f.v}\right] } | {{ init }} \rangle \nonumber = \sum_a
\langle - |\psi^R_a \rangle \langle \psi_a^L| {{ init }} \rangle \;
e^{-t\lambda_a(\mu)} 
\label{oop}
\end{equation}
If the spectrum has a gap,  the eigenvalue with the
lowest real part dominates, and we have that $G(\mu)=
\lambda_{{min}}(\mu)$ as $t \rightarrow \infty$.

Now, because $
\left[ H_K - \frac{\mu}{T} {\bf f.v}\right]$ and $\left[ H_K -
  \frac{1-\mu}{T} {\bf f.v}\right]^\dag $ are related
by a similarity transformation (\ref{che}), their spectra are the same.
Hence, we have that, to the extent that to leading order in the
 time $G(\mu)$ depends only on eigenvalues and not on eigenvectors:
\begin{equation}
G(\mu) = G(1-\mu)
\label{ft1}
\end{equation}
In this case,  (\ref{ft1}) is valid  only at large times.
Where can this fail? The problem in the large-time evaluation of
(\ref{oop}) arises if 
$ \langle \psi^L(\lambda_{min})| {{ init }} \rangle \sim 0$. At zero 
noise intensity this may well happen, because eigenvectors may in that 
case be completely localised.

\subsection{Gallavotti-Cohen theorem}

We shall not derive  here the Gallavotti-Cohen theorem, but just say
a few words. 
The Gallavotti Cohen Theorem is the  stationary fluctuation
theorem for a system in contact with  a 
{\em deterministic} Gaussian thermostat,
like the one we introduced in (\ref{Gau}) but with exactly
zero noise. It turns out that unlike the stochastic and
the transient case, which are essentially always valid,
the Gallavotti-Cohen result  breaks down in some systems,
and in most cases when  the forcing is very strong.
This is not a defect of the proof, but a reflection of a physical fact:
{\em the Fluctuation Relation in the deterministic case holds only
if the system has certain `ergodic' properties}.   

To have a perspective on this, one can consider approaching the 
deterministic case as a limit of the  noisy thermostatted case (\ref{Gau}).
One can reproduce the derivation above for this case~\cite{Kur}, and 
easily conclude that for every level of noise $\epsilon$ the theorem is valid,
 in the limit of large time windows $t \gg t_{min}(\epsilon)$, for some time $ t_{min}(\epsilon)$ .
The trouble comes from the fact that as $\epsilon \rightarrow 0$
it may happen that $t_{min}(\epsilon)$   diverges: 
in other words, the time $t_{min}(\epsilon)$ needed for the correct sampling
of fluctuations may become infinite in the deterministic limit.
The result of Gallavotti and Cohen proves that this is not the case 
for a class of very chaotic systems.
For this class their theorem  implies that
  $\lim_{\epsilon \rightarrow 0 } t_{min}(\epsilon)< \infty$.

The timescale  $t_{min}(\epsilon)$ has a clear physical meaning, which 
we just hint at here. A driven, deterministic,
 system has an {\em attractor} on the energy
surface (see sect 2.4).
 Winding back in time the dynamics defines a {\em repellor} which 
is stationary but unstable. In chaotic Hamiltonian (undriven) systems, attractor and
repellor are intertwined, they occupy the same region in phase-space. As the drive is turned
on, the attractor distribution focuses on a region the energy shell, as described in Sect. \ref{pp}. So
does the repellor, in a region that may be non-overlapping. 

 In the presence of weak noise, even if attractor and repellor are in principle separate,
the system  may occasionally switch from attractor to  repellor and
 back: this is exactly analogous  to the `coexistence' we found in lecture 5. The typical time to do this (\cite{Kur}, see also \cite{Bonetto})
 is precisely $t_{min}(\epsilon)$ for an otherwise ergodic system.
The condition for $t_{min}(\epsilon)$ to remain finite as the noise goes to zero
is then that attractor and repellor overlap sufficiently in
phase-space that it needs no noise to jump from one to the other.
 This is an extra condition the system must satisfy
in order that the Gallavotti-Cohen fluctuation theorem to holds.
{\em In a word, the  applicability of the 
Gallavotti-Cohen fluctuation theorem is, for a chaotic deterministic system,
a symptom  that attractor and repellor have not divorced under the effect of
forcing.}

\section{Jarzynski's equality}

Jarzynski's equality is a remarkable generalisation of the second principle.
Consider a system with an energy dependent upon a parameter (volume,
magnetic field, ...) which we denote $\alpha$.
We start from a Gibbs-Boltzmann  equilibrium corresponding to parameter
$\alpha_{initial}$ and then evolve while changing $\alpha(t)$ 
at arbitrary speed up to a time $t$ with $\alpha_{final}=\alpha(t)$.
 The equality is then:
\begin{equation}
e^{-\beta[ F(\alpha_{{final}})-F(\alpha_{{initial}} )]} = \langle e^{-\beta \; {\bf work} \; } 
\rangle_{ \alpha_{initial} \to \; \alpha_{final}}
\label{jarz}
\end{equation}
The average is over trajectories starting from equilibrium at $t=0$,
but otherwise arbitrary.  {\em Note the surprising appearance of
  $F(\alpha_{final})$, an equilibrium quantity, despite the fact that
  the system is not in equilibrium at time $t$}.
We shall  prove this result for a Langevin system with inertia.
It is valid independently of the friction
coefficient $\gamma$, indeed even if $\gamma=0$ and the system
is isolated. 

From (\ref{jarz}) we can go to the second principle, taking logarithms
on both sides and using Jensen's inequality $\langle e^A \rangle \geq
 e^{\langle A \rangle}$
\begin{equation}
F(\alpha_{{final}})-F(\alpha_{{initial}})\leq \; {\bf work}
\end{equation}
f
The time-dependent energy is:
\begin{equation}
{\mathcal{H}}_\alpha={\mathcal{H}}({\bf q,p},\alpha)= \sum_i
\frac{p_i^2}{2m} + V({\bf q},\alpha)
\end{equation}
If the parameter $\alpha$ depends on time, it does work:
\begin{eqnarray}
` & & {\mbox{force}} \times {\mbox{velocity}}= {\mbox energy \; change} -  {\bf work}
\nonumber \\  \int dt \; \sum_i \dot q_i \frac{\partial
  V}{\partial q_i} &=& \int dt \;
\left(\frac{d{\mathcal{H}}_\alpha}{dt} - \frac{\partial
  {\mathcal{H}}_\alpha}{\partial \alpha } \dot \alpha \right) =
   \left.  {\mathcal{H}}\right|^{{final}}_{{initial}} -
     \int dt \; \left(\frac{\partial {\mathcal{H}}_\alpha}{\partial
       t}\right) \nonumber 
\end{eqnarray}

\begin{figure}[htbp]
\begin{center}
\includegraphics[width=12cm]{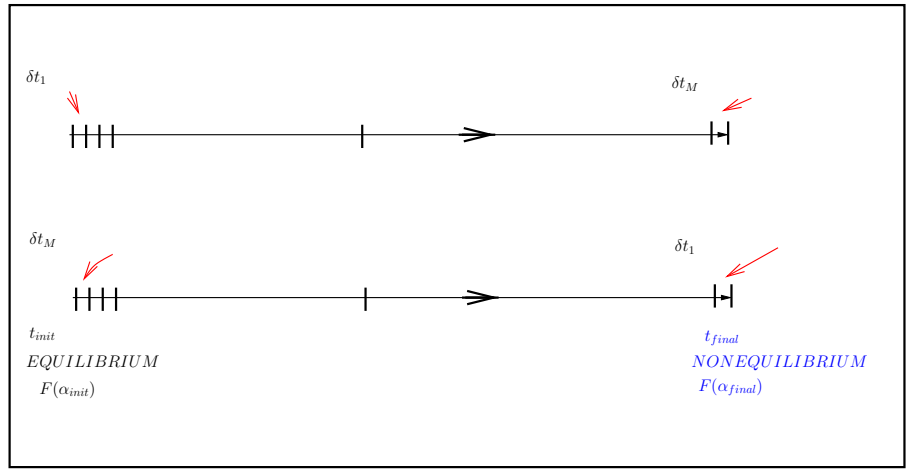}
\end{center}
\caption{Breaking the time-interval.}
\label{interval}
\end{figure}

We assume we start from the equilibrium configuration $|GB(\alpha_1)\rangle$
corresponding
to a given value  $\alpha_1$.
The total evolution over a time $t$ can be written by breaking the time
in short intervals  as in Fig. \ref{interval}.
\begin{equation}
U(t)= e^{-\delta t H_{\alpha_M}} ...  e^{-\delta t H_{\alpha_2}}
e^{-\delta t H_{\alpha_1 } } 
\end{equation}
Because $\langle - |H_{\alpha}=\langle - |$, we have
\begin{eqnarray}
1&=&\langle - |  e^{-\delta t H_{\alpha_M}} ...  e^{-\delta t H_{\alpha_2}}
e^{-\delta t H_{\alpha_1}}|GB(\alpha_1)\rangle
=\langle GB \; (\alpha_1)| e^{-\delta t H_{\alpha_1}^\dag}
e^{-\delta t H_{\alpha_2}^\dag} ...  e^{-\delta t H_{\alpha_M}^\dag}
|-\rangle \nonumber \\ &=&\frac{Z_{\alpha_M}}{Z_{\alpha_1}} \langle -
| e^{-\beta {\mathcal{H}}_{\alpha_1}} \; \; e^{-\delta t
  H_{\alpha_1}^\dag} e^{-\delta t H_{\alpha_2}^\dag} ...  e^{-\delta t
  H_{\alpha_M}^\dag} \;\; e^{\beta {\mathcal{H}}_{\alpha_M}} |GB \;
(\alpha_M)\rangle
\label{ooo}
\end{eqnarray}
As in all these theorems, we wish to introduce time-reversal.
We proceed as follows: we insert factors between every two
exponentials
\begin{equation}
\begin{array}{ccccc}
&& &{ \left(e^{\beta {\mathcal{H}}_{\alpha_2}} e^{-\beta
        {\mathcal{H}}_{\alpha_2}}\right)}& \\ e^{-\beta
    {\mathcal{H}}_{\alpha_1}} \; e^{-\delta t H_{\alpha_1}^\dag}
  &\uparrow& &\downarrow& e^{-\delta t H_{\alpha_2}^\dag} ...
  e^{-\delta t H_{\alpha_M}^\dag} \; e^{\beta
    {\mathcal{H}}_{\alpha_M}} \\ &{ \left(e^{\beta
      {\mathcal{H}}_{\alpha_1}} e^{-\beta
      {\mathcal{H}}_{\alpha_1}}\right)}& &&
\end{array}
\end{equation}
and use in each factor (\ref{hermik}):
\begin{equation}
 e^{-\beta {\mathcal{H}}_{\alpha_r}} e^{-\delta t H_{\alpha_r}^\dag}
 e^{\beta {\mathcal{H}}_{\alpha_r}}= \Pi e^{-\delta t H_{\alpha_r}}
 \Pi^{-1}
\end{equation}
Putting everything in (\ref{ooo}), using that the 
 $\Pi^2=1$ and that $ e^{-\delta t H_{\alpha_M}} \;
 |GB \; (\alpha_M)\rangle=|GB \; (\alpha_M)\rangle$
we get:
\begin{equation}
1 =\frac{Z_{\alpha_M}}{Z_{\alpha_1}} \langle - | e^{-\delta t
  H_{\alpha_1}} e^{-\beta
  ({\mathcal{H}}_{\alpha_1}-{\mathcal{H}}_{\alpha_2})} \; e^{-\delta t
  H_{\alpha_2}} e^{-\beta
  ({\mathcal{H}}_{\alpha_2}-{\mathcal{H}}_{\alpha_3})} ...  e^{-\delta
  t H_{\alpha_M}} |GB \; (\alpha_M)\rangle
\label{ooo1}
\end{equation}
Because ${\mathcal{H}}_{\alpha_r}-{\mathcal{H}}_{\alpha_(r+1)}  \propto \delta \alpha$, and using, as usual, that $e^{\delta t A} e^{\delta t B} \sim e^{\delta t
  (A+B)+O[(\delta(t))^2]}$ for small $\delta t$:
\begin{eqnarray}
1 &=&\frac{Z_{\alpha_M}}{Z_{\alpha_1}} \langle - | e^{-\delta t
  H_{\alpha_1}-\beta
  ({\mathcal{H}}_{\alpha_1}-{\mathcal{H}}_{\alpha_2})} \; e^{-\delta t
  H_{\alpha_2}-\beta
  ({\mathcal{H}}_{\alpha_2}-{\mathcal{H}}_{\alpha_3})} ...  e^{-\delta
  t H_{\alpha_M}-\beta
  ({\mathcal{H}}_{\alpha_{M-1}}-{\mathcal{H}}_{\alpha_M})} |GB \;
(\alpha_M)\rangle\nonumber \\ &=&\frac{Z_{\alpha_M}}{Z_{\alpha_1}}
\langle - | e^{-\delta t H_{\alpha_1}-\beta \frac{\partial
    {\mathcal{H}}_{\alpha_1}}{\partial \alpha} \delta \alpha } \; e^{-\delta t
  H_{\alpha_2}-\beta \frac{\partial {\mathcal{H}}_{\alpha_2}}{\partial
    \alpha} \delta \alpha } ...  e^{-\delta t H_{\alpha_M}-\beta \frac{\partial
    {\mathcal{H}}_{\alpha_{M-1}}}{\partial \alpha}\delta \alpha } |GB \;
(\alpha_M)\rangle
\label{ooo2}
\end{eqnarray}
which, by simple comparison means that
\begin{equation}
1= e^{\beta(    F_{\alpha_M}-F_{\alpha_1})} \langle e^{-\beta \int dt \;
\frac{\partial {\mathcal{H}}}{\partial t}} \rangle
\end{equation}
where the average is over trajectories starting from initial points
chosen with the  equilibrium distribution at $(\alpha_M)$ and ending anywhere
where evolution and noise takes them. 

We finally get
\begin{equation}
e^{-\beta( F_{{final}}-F_{{initial}} )} = \langle e^{-\beta \int dt \;
  \frac{\partial {\mathcal{H}}}{\partial t} } \rangle_{
  \alpha_{initial} \to \; any}
\end{equation}
The average is over trajectories starting from equilibrium at
$\alpha_{initial}$ and ending anywhere that the dynamics takes them.
The interpretation of  $work \; = \;\int dt \; \frac{\partial
{\mathcal{H}}}{\partial t}$  has generated controversy~\cite{RubiLuca}, 
it is the work done by the system on the 
external sources (e.g. pistons, etc) of the `fields' $\alpha$.
 Here again, subtly different  equalities can be
obtained depending on the precise expression used for work, see 
\cite{Jarzynski-difference}.

\section{A paradox}

The   Jarzynski equality has been criticised~\cite{Crooks_Jarzynski} on the basis that  
it it is supposed to fail
 for a process of free expansion. This paradox was completely resolved
by Crooks and Jarzynski. The resolution is in itself instructive,
because it highlights the role of rare fluctuations.
The argument goes as follows: if we open the tap of a bottle with gas, 
letting it freely expand into an empty room, there is no work done,
and yet there is a change in the free energies before and after:
Jarzynski's equality must be violated. 

\begin{figure}[htbp]
\begin{center}
\includegraphics[width=12cm]{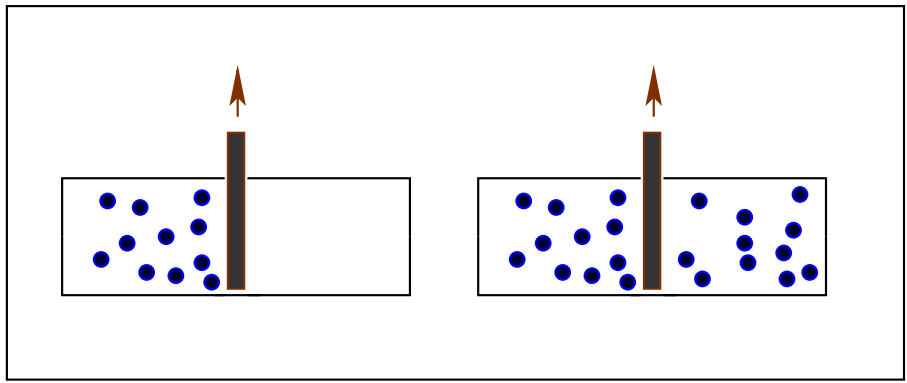}
\end{center}
\caption{Sliding wall. Left and right: nonequilibrium and equilibrium initial
conditions. }
\label{sliding}
\end{figure}

\begin{figure}[htbp]
\begin{center}
\includegraphics[width=12cm]{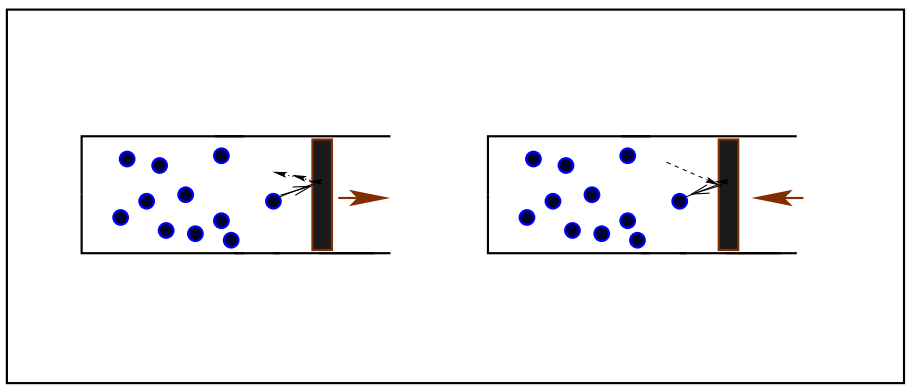}
\end{center}
\caption{Pulling back a piston fast. The process and its time-reversed.
}
\label{Pulling}
\end{figure}

Let us first distinguish two ways of making a free expansion: with a 
sliding
wall (Fig. \ref{sliding}) or with a 
rapidly receding piston (Fig. \ref{Pulling}).
In the case of the sliding wall we see immediately where is the catch:
we are supposed to start with an equilibrium initial condition, but this 
requires that the right-hand compartment in Fig. \ref{sliding}
be also full. If such is the case,
when we slide the wall open there is neither work done nor free-energy 
change.
 
This seems like a cheat, because we could have a solid piston receding 
infinitely fast as in Fig \ref{Pulling}, and in that case no need of an equilibrated right hand 
compartment.
 Following Crooks and Jarzynski, suppose then that  the
 piston is pulled backwards a huge velocity $v$. If $v\gg (kT)^{1/2}$, it is
 highly unlikely that any gas particle will hit the piston as it
 recedes, so in almost  any run of the experiment no work is
 done. There are, however,  very rare realisations of the experiment in
 which an unusually energetic particle catches up with the piston,
 bounces, and looses most of its velocity in the process.  This can be
 best seen if one considers the time-reversed process (Fig \ref{Pulling})
in which a
 rapidly incoming piston hits a slow moving particle.  It turns
 out~\cite{Crooks_Jarzynski} that the rare realizations in which this strong
 `particle cooling' happens suffice to account for the exponential of
 the free energy difference.

This example beautifully illustrates the role of rare fluctuations
in Jarzynski's equality, and the extent to which what dominates the equation
can be an extremely atypical process that violates our intuition,
which is designed to apply to probable events.

\section{Experimental work}

A serious account of both the Fluctuation theorem and Jarzynski's equality
should discuss the, by now quite considerable,  experimental work.
We shall not do this here, but just refer
the interested reader to the  References ~\cite{ritort}. 

Let us make however a few remarks, from a theoretician's perspective.
The fluctuation theorem is, as its name suggests, a theorem. As such
it need not be tested experimentally. The question is, of course, to what
extent a specific physical system satisfies the hypotheses. 
Clearly, no real system has a Gaussian thermostat. Even a Langevin thermostat
is an idealisation --
a question that has been brought up in the context of Jarzynski's equality, 
(see below).
The usual way out is to say that the thermostat is `far away'
and then its nature becomes irrelevant. This would normally sound extremely
reasonable, but bear in mind that, as the previous section shows, intuition
may be  misleading when applied to large deviations.
At any rate, if the nature of the thermostat is indeed not important,
then we might as well suppose it is a stochastic one, so that the fluctuation
theorem holds without any assumption related to ergodicity.

Consider for example the Lyon experiment~\cite{Lyon} where a liquid
is enclosed  between
two horizontal plates, the lower at higher temperature 
than the upper one $T_d>T_u$.
Heat is transmitted by (Rayleigh-Benard) convection. The fluctuation
theorem establishes a relation between having a heat flux ${\mathcal J}$ in one sense
and in the other, $P({\mathcal J})/P(-{\mathcal J}) \sim e^{-(\beta_u-\beta_d){\mathcal J} t}$ 
\cite{chain}. Surely, one argues, the top and bottom plates can be considered
to be in contact with a good Langevinian bath,  the nature 
of Rayleigh-Benard convection cannot depend on that.
If this assumption is correct, then the validity of the fluctuation theorem 
is  not in doubt, as no ergodicity properties are required of the system when there is
a Langevin bath.
 But in fact, the experiment does not find a fluctuation
relation with the true top and bottom temperatures (and indeed, with such temperatures
the violations of the second law would be unobservable). Why is that?.
   It would seem -- but this needs further elucidation --
that what the experiment is in fact testing is a 
pre-asymptotic result, because the time-window is not long enough.
One can of course ask if at this pre-asymptotic level there is 
for some other reason {\em another}
fluctuation relation, with a higher effective temperature, but this requires
a different theory~\cite{Bonetto}.

Let us now turn very briefly to the Jarzynski equality. In this case,
experimental work~\cite{ritort} has been made not so much to test
the equality, as in the case of the fluctuation theorem, but to {\em
  use} it to evaluate free-energy differences with fast, out of
equilibrium, measurements.  For example, two RNA strands are unzipped
by pulling with optical tweezers, and the work that this costs is
measured~\cite{bustamante}. The experiment is repeated many times, and from the work
distribution the equilibrium free-energy difference is measured.  If
we move too fast, it is the rarer and rarer runs that dominate, so the
experiment has to be repeated many times. Here again, the example of
the piston in the previous section is very illuminating: if the piston
recedes very fast, only in very rare repetitions of the experiment
will a fast particle catch up with it and be stopped: but this event
dominates the average in Jarzynski's equality!.

Let us say a
 final word concerning Jarzynski's equality in the presence of a (water) bath.
It has been argued ~\cite{Cohen_Mauzerall} that the water can not be considered as a 
Langevinian -- or in fact as any other equilibrium --
 bath, while the system is evolving
fast. This is indeed so, but the problem is solved~\cite{Jarzynski} by
'moving the bath away', i.e. by including the surrounding water as
  part of the system.

\end{document}